\documentclass[amssymb,aps,prd,floatfix]{revtex4-2}
\usepackage[intlimits]{amsmath}
\usepackage{graphicx}
\usepackage{tensor}
\usepackage[dvipsnames]{xcolor}
\usepackage{hyperref}
\usepackage{epstopdf}
\usepackage{epsfig}
\usepackage{color}
\usepackage{subfigure,url}
\begin{document}

\title{Finite collisionless accretion disks in the Kerr spacetime}

\author{Ghafran Khan}
\affiliation{Szko{\l}a Doktorska Nauk \'{S}cis{\l}ych i Przyrodniczych, Uniwersytet Jagiello\'{n}ski}
\affiliation{Instytut Fizyki Teoretycznej, Uniwersytet Jagiello\'{n}ski, \L ojasiewicza 11, 30-348 Krak\'ow, Poland}
\author{Patryk Mach}
\affiliation{Instytut Fizyki Teoretycznej, Uniwersytet Jagiello\'{n}ski, \L ojasiewicza 11, 30-348 Krak\'ow, Poland}

\begin{abstract}
We construct general-relativistic kinetic models of stationary finite accretion disks in the Kerr spacetime. Our analysis generalizes a previous model of razor-thin accretion disks of collisionless gas in the Kerr spacetime, extending to infinity in the equatorial plane. We investigate monoenergetic configurations, as well as models characterized by Maxwell-J\"{u}ttner type distributions at the outer edge of the disk. In both cases, we consider particles moving along orbits reaching the outer boundary of the disk, either plunging into the black hole, or scattered by the centrifugal barrier. For Maxwell-J\"{u}ttner models, the location of the outer disk boundary affects the mass and angular momentum accretion rates at a similar order of magnitude as the temperature parameter in the assumed Maxwell-J\"{u}ttner distribution.
\end{abstract}

\maketitle

\section{Introduction}

General-relativistic kinetic models of matter---as opposed to more familiar hydrodynamical or magnetohydrodynamical ones---apply to astrophysical situations in which collisions between gas particles are rare and the assumption of the local thermodynamic equilibrium is no longer valid. This may happen for a plasma in a strong gravitational field \cite{Parfrey,Bransgrove,Crinquand,Galishnikova}, for dark matter particles \cite{Dominguez2017}, or simply for galaxies or stellar clusters described as a gas of stars \cite{Shapiro1993a,Shapiro1993b}.

Recently, a kinetic approach was used in general-relativistic simulations of magnetized plasma accreting on the Kerr black hole, allowing for a direct comparison with magnetohydrodynamical models \cite{Parfrey,Bransgrove,Crinquand,Galishnikova}. At the same time, much effort has been devoted to understand the behavior of stationary models, even without magnetic fields (see, for instance, a recent work \cite{Shapiro2023}). A modern approach to a general-relativistic kinetic counterpart of the Bondi accretion---steady, spherically symmetric accretion flow onto a central gravitating object---was developed by Rioseco and Sarbach in \cite{Olivier1,Olivier2} for a Schwarzschild black hole. An analogous model with a moving Schwarzschild black hole (the so-called Bondi-Hoyle-Lyttleton accretion) was analyzed in \cite{pmao1,pmao2,pmao3}. The analysis of \cite{Olivier1,Olivier2} was repeated for the Reissner-Nordstr\"{o}m black hole in \cite{cieslik}, but also for more exotic spacetimes \cite{exotic1,exotic2,exotic3}.

Analytic studies of matter around black holes performed within the framework of the general-relativistic kinetic theory include also recent disk models \cite{Gabarrete2023}, works on phase-space mixing \cite{Rioseco2018,Rioseco2024} or a preliminary work dealing with Einstein-Vlasov (self-gravitating) systems in the presence of black holes \cite{Andreasson2021}.

This paper is a sequel to a previous study of accretion of a collisionless gas confined to the equatorial plane in the Kerr spacetime, intended to serve as a toy model of stationary dark matter accretion onto a rotating black hole \cite{CMO2022} (see also \cite{MO2023,Li2023}). The conditions assumed in \cite{CMO2022} mimicked to some extent those of the Bondi accretion---the gas was assumed to extend to infinity in the equatorial plane. In a physical situation, the accretion starts from a finite distance. Unfortunately, taking this fact into account adds a complication to the model. In the analysis presented in this paper, the boundary conditions are assumed at a finite radius at which the matter is added to the system. This implies that particles movig along bounded trajectories must be taken into account whenever such trajectories can reach the outer boundary. Another difficulty is related to the boundary distribution, which, in contrast to the case of infinite disks, is not defined on a flat spacetime region. Both complications appear also in the spherically symmetric model of accretion onto the Schwarzschild black hole occurring from a sphere of a finite radius \cite{olivierfinite}. In this paper, we use an approach similar to the one discussed in \cite{olivierfinite}, adapted to the equatorial motion in the Kerr geometry.

We investigate two models of the gas consisting of spinless same-mass particles. In the first one, all of the gas particles are characterized by the same energy. Although this assumption is clearly nonphysical, it allows one to demonstrate the effects associated with the phase-space geometry associated with the Kerr spacetime. The second model is obtained assuming a Maxwell-J\"{u}ttner type distribution at the outer disk boundary. In both cases, we provide exact expressions for the particle current surface density as well as for accretion rates of the mass, the energy, and the angular momentum.

The order of this paper is as follows. In Section \ref{sec:Preliminaries} we discuss all relevant elements of our model. In particular, we introduce horizon-penetrating Kerr coordinates, describe the geodesic motion confined to the equatorial plane of the Kerr spacetime, and define main quantities related to the kinetic description of the gas. The main novel element is related to the analysis of the radial motion of particles originating at a fixed finite distance from the central black hole. This analysis gives rise to a description of the ranges in the phase space available for the motion of individual gas particles. In Section \ref{sec:accretionmodel} we describe the general accretion model of this paper. This part is then expanded into two separate sections. In Section \ref{sec:monoenergetic} we deal with monoenergetic particles. Section \ref{sec:maxwelljuttner} is devoted to the accretion model with the gas obeying a Maxwell-J\"{u}ttner type distribution at the outer edge of the disk. Section \ref{sec:summary} contains a summary.

We work in standard geometric units with $c = G = 1$, where $c$ denotes the speed of light and $G$ is the gravitational constant. The metric signature is $(-,+,+,+)$.

\section{Preliminaries}
\label{sec:Preliminaries}

\subsection{Geodesic motion in the Kerr spacetime}

In Boyer-Lindquist coordinates $(t^\prime,r^\prime,\vartheta^\prime,\varphi^\prime)$ the Kerr metric can be written as \cite{BoyerLindquist1967}
\begin{equation}
\label{kerr4d}
    g = -\left( 1 - \frac{2Mr^\prime}{\rho^2} \right) {dt^\prime}^2 - \frac{4 M a r^\prime \sin^2 \vartheta^\prime}{\rho^2} dt^\prime d\varphi^\prime + \frac{\rho^2}{\Delta} {dr^\prime}^2 + \rho^2 {d\vartheta^\prime}^2 + \left( {r^\prime}^2 + a^2 + \frac{2 M a^2 r^\prime \sin^2 \theta^\prime}{\rho^2} \right) \sin^2 \vartheta^\prime {d\varphi^\prime}^2,
\end{equation}
where
\begin{subequations}
\begin{eqnarray}
\Delta & = & {r^\prime}^2 - 2 M r^\prime + a^2, \\
\rho^2 & = & {r^\prime}^2 + a^2 \cos^2 \vartheta^\prime.
\end{eqnarray}
\end{subequations}
Since we are interested in the accretion problem, it is more appropriate to work in horizon-penetrating coordinates. One of simplest coordinate systems of this type, denoted further as $(t,r,\vartheta,\varphi)$, can be obtained by setting
\begin{equation}
t^\prime = t - 2M\int^r \frac{r dr}{\Delta},\qquad
\varphi^\prime = \varphi - a\int^r\frac{dr}{\Delta}, \quad r^\prime = r, \quad \vartheta^\prime = \vartheta.
\end{equation}
We will refer to $(t,r,\vartheta,\varphi)$ as the horizon-penetrating Kerr coordinates. In the astrophysical literature this coordinate system is also known as Kerr-Schild coordinates \cite{rezzolla}. We would like to stress that a nearly identical coordinate system was used in the original paper by Kerr \cite{Kerr}, save for a change $t = u - r$ and a different convention regarding the spin parameter $a \to -a$.

In horizon-penetrating Kerr coordinates the Kerr metric can be written as
\begin{equation}
g = -dt^2 +dr^2 -2a\sin^2\vartheta dr d\varphi 
 + \left( r^2 + a^2 \right)\sin^2 \vartheta d\varphi ^2
 + \rho^2 d\vartheta ^2 
 + \frac{2M r}{\rho^2}  \left(dt + dr - a\sin ^2 \vartheta d\varphi \right)^2.
\label{Eq:Kerr}
\end{equation}
The inverse metric has the form
\begin{equation}
(g^{\mu\nu}) = \frac{1}{\rho^2}\left( \begin{array}{cccc}
 -(\rho^2 + 2M r) & 2Mr & 0 & 0 \\
 2Mr & \Delta & 0 & a \\
 0 & 0 & 1 & 0 \\
 0 & a & 0 & \sin^{-2}\vartheta
\end{array} \right),
\label{Eq:KerrInverse}
\end{equation}
and its determinant reads
\begin{equation}
\mathrm{det} \, g^{\mu \nu} = - \frac{1}{\rho^4 \sin^2 \vartheta}.
\end{equation}

Time slices of the event and Cauchy horizons are coordinate spheres with radii $r_\pm = M \pm \sqrt{M^2 - a^2}$, roots of the equation $\Delta(r) = 0$. The Kerr metric admits the following two Killing vectors, which in horizon-penetrating Kerr coordinates can be expressed as
\begin{equation}
\label{killing}
k^\mu = (1,0,0,0), \quad \chi^\mu = (0,0,0,1).
\end{equation}

Timelike geodesics can be described by Hamilton's equations of the form
\begin{equation}
\label{hamiltoneqs}
\frac{dx^\mu}{d \tau} = \frac{\partial \mathcal H}{\partial p_\mu}, \quad \frac{dp_\nu}{d \tau} = \frac{\partial \mathcal H}{\partial x^\nu},
\end{equation}
with the Hamiltonian
\begin{equation}
\mathcal H = \frac{1}{2}g^{\mu \nu}(x^\alpha) p_\mu p_\nu = \frac{1}{2 \rho^2} \left[ -(\rho^2 + 2 M r) p_t^2 + 4 M r p_t p_r + \Delta p_r^2 + p_\vartheta^2 + 2 a p_r p_\varphi + \frac{1}{\sin^2 \vartheta} p_\varphi^2 \right].
\end{equation}
System (\ref{hamiltoneqs}) admits the following four constants of motion \cite{Carter1968}:
\begin{subequations}
\begin{eqnarray}
    m^2 & = & - g^{\mu \nu} p_\mu p_\nu, \\
    E & = & - p_t, \\
    l_z & = & p_\varphi, \\
    l^2 & = & p_\vartheta^2 + \left( \frac{p_\varphi}{\sin \vartheta} + a \sin \vartheta  p_t \right)^2 + m^2 a^2 \cos^2 \vartheta.
\end{eqnarray}
\end{subequations}
As a consequence the equations of motion can be separated as follows
\begin{subequations}
\begin{eqnarray}
    p_t & = & -E, \\
    (\Delta p_r - 2 M E r + a l_z)^2 & = & R(r), \\
    p_\vartheta^2 & = & l^2 - \left( \frac{p_\varphi}{\sin \vartheta} + a \sin \vartheta  p_t \right)^2 - m^2 a^2 \cos^2 \vartheta, \label{ptheta} \\
    p_\varphi & = & l_z,
\end{eqnarray}
\end{subequations}
where
\begin{equation}
R(r) = \left[ E (r^2 + a^2) - a l_z \right]^2 - \Delta (l^2 + m^2 r^2).
\end{equation}
There is a large body of literature devoted to solutions of geodesic equations (\ref{hamiltoneqs}). Textbook introductions can be found in \cite{Chandrasekhar,ONeill}. Solutions of geodesic equations (\ref{hamiltoneqs}) in horizon-penetrating Kerr coordinates were recently studied in \cite{Bakun2024} within a Weierstrass framework introduced in \cite{CieslikHackmannMach2023}.

\subsection{Geodesic motion at the equatorial plane}

Restricting ourselves to the equatorial plane, we set $p_\theta = 0$, $\theta = \pi/2$. Equation (\ref{ptheta}) then gives
\begin{equation}
l^2 = (p_\varphi + a p_t)^2 = (l_z - a E)^2.
\end{equation}
We will adopt a convention with $l \ge 0$ and
\begin{equation}
l_z = \epsilon_\sigma l + aE,
\end{equation}
where $\epsilon_\sigma = \pm 1$.
For the orbit restricted to the equatorial plane the geodesic equations can be written as
\begin{subequations}
\begin{eqnarray}
\frac{dt}{d\tau} & = & p^t = \frac{(r + 2M) E + 2 M p_r}{r}, \\
\frac{dr}{d \tau} & = & p^r = \frac{\Delta p_r - 2 M E r + a l_z}{r^2}  = \epsilon_r \frac{\sqrt{R(r)}}{r^2}, \\
\frac{d\varphi}{d \tau} & = & p^\varphi = \frac{l_z + a p_r}{r^2},
\end{eqnarray}
\end{subequations}
where $\epsilon_r = \pm 1$ corresponds to the direction of the radial motion.

The geodesic motion confined to the equatorial plane can also be analyzed basing on the 2 + 1 dimensional metric $\gamma_{\mu \nu}$ induced in the equatorial plane \cite{CMO2022}. It reads (in the Kerr-Schild form)
\begin{equation}
\gamma = -dt^2 + dr^2 - 2 a dr d\varphi 
 + \left( r^2 + a^2 \right) d\varphi ^2
 + \frac{2M}{r}  \left(dt + dr - a d\varphi \right)^2.
\end{equation}
The inverse metric can be written as
\begin{equation}
(\gamma^{\mu\nu}) = \frac{1}{r^2}\left( \begin{array}{ccc}
 -(r^2 + 2M r) & 2Mr & 0 \\
 2Mr & \Delta & a \\
 0 & a & 1
\end{array} \right),
\end{equation}
where the coordinates are ordered as $(t,r,\varphi)$. It is easy to check that $\sqrt{- \mathrm{det}\, \gamma^{\mu \nu}} = 1/r$.

\subsection{Vlasov equation}
\label{sec:vlasov}

In the next four subsections, we review some key elements of the general-relativistic kinetic theory, restricting ourselves to the collisionless case. Excellent general introductions to the topic can be found in \cite{Sarbach2014,intro} (see also \cite{Andreasson2011} for an introduction to Vlasov-Einstein systems). We will mainly focus on the differences between a description of the general three-dimensional case and a two-dimensional configuration of a gas confined to the equatorial plane of the Kerr spacetime. This aspect has been discussed to some extent in \cite{CMO2022}.

Let $(\mathcal{M},g)$ denote the spacetime manifold. The cotangent bundle on $(\mathcal{M},g)$ is defined as
\begin{equation}
T^\ast \mathcal M = \{ (x,p) \colon x \in \mathcal{M}, p \in T_x^\ast \mathcal M \},  \end{equation}
where $T_x^\ast \mathcal M$ denotes the cotangent space to $\mathcal{M}$ at the point $x$. The Vlasov gas---a collection of individual particles---can be described in terms of the one-particle distribution function $\mathcal F \colon U \to [0,+\infty)$, where $U \subseteq T^\ast \mathcal M$. It can be given the following statistical interpretation. Let $S$ be a three-dimensional spacelike surface in $\mathcal M$. Define $\mathcal N[S]$ as an averaged number of particle trajectories in $U$, whose projections on $\mathcal M$ intersect the surface $S$. One can show that (cf.\ \cite{intro}, Eqs. (94))
\begin{equation}
\mathcal N[S] = - \int_S \left[ \int_{P_x^+} \mathcal F(x,p) p_\mu s^\mu \mathrm{dvol}_x(p) \right] \eta_S,
\end{equation}
where $s$ is a unit vector normal to $S$, $\eta_S$ denotes the three-dimensional volume element induced on $S$,
\begin{equation}
P_x^+ = \{ p \in T_x^\ast \mathcal M \colon g^{\mu \nu} p_\mu p_\nu < 0, \, p \text{ is future directed} \},
\end{equation}
and $\mathrm{dvol}_x(p)$ denotes the volume element in $P_x^+$, given by
\begin{equation}
\mathrm{dvol}_x(p) = \sqrt{- \mathrm{det} g^{\mu \nu}(x)} dp_0 dp_2 dp_2 dp_3.
\end{equation}
Defining the so-called particle current density
\begin{equation}
\mathcal J_\mu(x) = \int_{P_x^+} \mathcal F(x,p) p_\mu \mathrm{dvol}_x(p),
\end{equation}
one can write the expression for $\mathcal N[S]$ as
\begin{equation}
\mathcal N[S] = - \int_S \mathcal J_\mu s^\mu \eta_S.
\end{equation}
The energy-momentum tensor of the Vlasov gas is defined as
\begin{equation}
\mathcal T_{\mu \nu} (x) = \int_{P_x^+} \mathcal F(x,p) p_\mu p_\nu \mathrm{dvol}_x(p). 
\end{equation}

If collisions between individual gas particles are neglected, the particles move along geodesics, and the distribution function $\mathcal F$ should remain constant along geodesics. This amounts to the following condition
\begin{eqnarray}
\label{vlasov}
    \frac{d \mathcal F}{d \tau}=\frac{\partial \mathcal F}{\partial x^\mu} \frac{d x^\mu}{d \tau}+\frac{\partial \mathcal F}{\partial p_\nu} \frac{d p_\nu}{d \tau}= \frac{\partial \mathcal F}{\partial x^\mu} \frac{\partial \mathcal H}{\partial p_\mu}-\frac{\partial \mathcal F}{\partial p_\nu} \frac{\partial \mathcal H}{\partial x^\nu} = \{\mathcal H, \mathcal F \} = 0,
\end{eqnarray}
where $\{._,.\}$ denotes the Poisson bracket. Equation (\ref{vlasov}) is referred to as the Vlasov equation.

\subsection{Vlasov equation at the equatorial plane}
\label{sec:vlasovequatorial}

Let us return to the motion in the Kerr spacetime. For the motion confined to the equatorial plane $\vartheta = \pi/2$, it is convenient to define the surface distribution function $F(t,r,\varphi;p_\alpha)$ and surface densities $J_\mu(t,r,\varphi)$ and $T_{\mu \nu}(t,r,\varphi)$ by
\begin{equation}
\mathcal F(t,r,\vartheta,\varphi;p_\alpha) = \delta(z) F(t,r,\varphi;p_\alpha), \quad \mathcal J_\mu (t,r,\vartheta,\varphi) = \delta(z) J_\mu(t,r,\varphi), \quad \mathcal T_{\mu \nu} (t,r,\vartheta,\varphi) = \delta(z) T_{\mu \nu} (t,r,\varphi),
\end{equation}
where the $z$ coordinate is chosen in such a way that the vector $\partial_z$ is of unit length and is normal to the equatorial plane $\vartheta = \pi/2$. Note that the coordinate $z$ must only be defined in the vicinity of the equatorial plane. Consequently, it suffices to take the relation $z = r \cos \vartheta$. This gives $\delta(z) = \delta(\vartheta-\pi/2)/r$. With these definitions,
\begin{equation}
\label{jmutmunusurface}
J_\mu(t,r,\varphi) = \int_{P_x^+} F(t,r,\varphi;p_\alpha) p_\mu \mathrm{dvol}_x(p), \quad T_{\mu \nu} (t,r,\varphi) = \int_{P_x^+} F(t,r,\varphi;p_\alpha) p_\mu p_\nu \mathrm{dvol}_x(p).
\end{equation}

The consistency condition for the motion confined to the equatorial plane is that
\begin{equation}
F(t,r,\varphi,p_t,p_r,p_\vartheta,p_\varphi) = \delta(p_z) f(t,r,\varphi,p_t,p_r,p_\varphi), \end{equation}
where $p_z$ is the momentum component associated with the coordinate $z$. It ensures that the particles have no momentum perpendicular to the equatorial plane, and consequently remain at $\vartheta = \pi/2$. Since at the equatorial plane $p_z = - p_\vartheta/r$, one has $\delta(p_z) = r \delta(p_\vartheta)$. Note that at the equatorial plane the momentum space volume element reduces to
\begin{equation}
\mathrm{dvol}_x(p) = \frac{1}{\rho^2 \sin \vartheta} dp_t dp_r dp_\vartheta dp_\varphi = \frac{1}{r^2}  dp_t dp_r dp_\vartheta dp_\varphi. \end{equation}
Taking the expression $F(t,r,\varphi,p_t,p_r,p_\vartheta,p_\varphi) \mathrm{dvol}_x(p)$ restricted to the equatorial plane and integrating over $p_\vartheta$, we get
\begin{eqnarray}
\int F(t,r,\varphi,p_t,p_r,p_\vartheta,p_\varphi) \mathrm{dvol}_x(p) & = & \int  \delta(p_\vartheta) f(t,r,\varphi,p_t,p_r,p_\varphi) \frac{dp_t dp_r dp_\vartheta dp_\varphi}{r}  \nonumber \\
& = & f(t,r,\varphi,p_t,p_r,p_\varphi) \sqrt{-\mathrm{det} \, \gamma^{\mu \nu}} dp_t dp_r dp_\varphi.
\end{eqnarray}

It is easy to show that the distribution function $f = f(t,r,\varphi,p_t,p_r,p_\varphi)$ satisfies the Vlasov equation of the form
\begin{equation}
\label{vlasov3d}
\frac{d f}{d \tau}=\frac{\partial f}{\partial x^\mu} \frac{d x^\mu}{d \tau}+\frac{\partial f}{\partial p_\nu} \frac{d p_\nu}{d \tau}= \frac{\partial  f}{\partial x^\mu} \frac{\partial H}{\partial p_\mu}-\frac{\partial f}{\partial p_\nu} \frac{\partial H}{\partial x^\nu} = 0,
\end{equation}
where
\begin{equation}
H = \frac{1}{2} \gamma^{\mu \nu} (x^\alpha) p_\mu p_\nu
\end{equation}
and $x^\mu = (t,r,\varphi)$, $p_\nu = (p_t,p_r,p_\varphi)$.

The form of Eqs.\ (\ref{vlasov}) [or (\ref{vlasov3d})] implies that a distribution function that depends on the phase-space variables $(x^\mu,p_\nu)$ via constants of the geodesic motion
\begin{equation}
    I_i = I_i(x^\mu,p_\nu), 
    \quad \{ \mathcal H, I_i \} = 0, \quad i = 1, \dots, s,
\end{equation}
i.e., $\mathcal F = \mathcal F(I_1(x^\mu,p_\nu), \dots, I_s(x^\mu,p_\nu))$, satisfies the Vlasov equation. For the general Vlasov equation on the Kerr spacetime this implies that any function of the form $\mathcal F = \mathcal F(m,E,l_z,l)$ would satisfy Eq.\ (\ref{vlasov}). Similarly, any distribution function $f = f(m, E, l)$ or $f = f(m, E, l_z)$ would satisfy Eq.\ (\ref{vlasov3d}). There is an elegant formal generalization of this result that can be obtained in terms of suitably chosen action-angle variables (see, e.g., \cite{Olivier1,Sarbach2014}). We will discuss it shortly for the three-dimensional case of the motion restricted to the equatorial plane. Consider a canonical transformation $(t,r,\varphi,p_t,p_r,p_\varphi) \mapsto (Q^0,Q^1,Q^2,P_1,P_2,P_3)$. The new momenta $P_0$, $P_1$, and $P_3$ are set to be equal to constants of motion
\begin{eqnarray}
    P_0 = m, \quad P_1 = E, \quad P_2 = l_z.
\end{eqnarray}
The conjugate variables $Q^1$, $Q^2$, and $Q^3$ are defined as
\begin{equation}
Q^0 = \frac{\partial W}{\partial m}, \quad Q^1 = \frac{\partial W}{\partial E}, \quad Q^2=\frac{\partial W}{\partial l_z},
\end{equation}
where the generating function $W$ reads
\begin{eqnarray}
W(t,r,\varphi,m,E,l_z) & = &  \int p_t dt + \int p_r dr + \int p_\varphi d\varphi = -E t + l_z \varphi + \int p_r dr \nonumber \\
& = & -E t + l_z \varphi + \int \frac{\epsilon_r \sqrt{R(r)} + 2 M E r - a l_z}{\Delta} dr,
\end{eqnarray}
and
\begin{equation}
    R(r) =  \left[ E (r^2 + a^2) - a l_z \right]^2 - \Delta \left[ (l_z - a E)^2 + m^2 r^2 \right]
\end{equation}
for the motion confined to the equatorial plane. In explicit terms, $(Q^0,Q^1,Q^2)$ can be written as
\begin{subequations}
\begin{eqnarray}
    Q^0 & = & m \int \frac{\epsilon_r r^2}{\sqrt{R(r)}} dr, \\
    Q^1 & = & -t + \int \left\{ \frac{2 M r}{\Delta} + \frac{\epsilon_r}{\Delta \sqrt{R(r)}} \left[ E r^2 (r^2 + a^2) - 2 M r a (l_z - a E) \right] \right\} dr, \\
    Q^2 & = & \varphi + \int \left\{ -\frac{a}{\Delta} + \frac{\epsilon_r}{\Delta \sqrt{R(r)}} \left[ - r^2 l_z + 2 M r (l_z - a E) \right] \right\} dr.
\end{eqnarray}
\end{subequations}
The Vlasov equation (\ref{vlasov3d}), expressed in terms of the Poisson bracket, is covariant with respect to this canonical transformation. Since $H = - \frac{1}{2}m^2 = - \frac{1}{2} P_0^2$, it takes, in new variables, the form
\begin{equation}
- P_0 \frac{\partial f}{\partial Q^0} = 0.\end{equation}
Consequently, any function of the form
\begin{equation}
f = f(Q^1,Q^2,m,E,l_z)
\end{equation}
satisfies Eq.\ (\ref{vlasov3d}). Further restrictions on the form of $f$ may follow from symmetry requirements.

A Killing vector field on $(\mathcal M,g)$ expressed as
\begin{equation}
\xi_x = \xi^\mu(x) \left.  \frac{\partial}{\partial x^\mu}  \right|_x
\end{equation}
can be extended to the contangent bundle $T^\ast \mathcal M$ by defining the so-called lift
\begin{equation}
\hat \xi_{(x,p)} =  \xi^\mu(x) \left.  \frac{\partial}{\partial x^\mu}  \right|_{(x,p)} - p_\alpha \frac{\partial \xi^\alpha}{\partial x^\mu}(x) \left. \frac{\partial}{\partial p_\mu} \right|_{(x,p)}.
\end{equation}
The lifts defined in this way satisfy Killing equations with respect to the so-called Sasaki metric---a natural metric on $T^\ast \mathcal M$ \cite{intro}. The lifts of the two Killing vectors $k = \partial_t$ and $\chi = \partial_\varphi$, admitted by the Kerr spacetime, read
\begin{equation}
\hat k = \frac{\partial}{\partial t}, \quad \hat \chi = \frac{\partial}{\partial \varphi}.
\end{equation}
They can be expressed, in new coordinates $(Q^\mu,P_\nu)$ as
\begin{subequations}
\begin{eqnarray}
\hat k & = & \frac{\partial}{\partial t} = \frac{\partial Q^\mu}{\partial t} \frac{\partial}{\partial Q^\mu} + \frac{\partial P_\nu}{\partial t} \frac{\partial}{\partial P_\nu} = - \frac{\partial}{\partial Q^1}, \\
\hat \chi & = & \frac{\partial}{\partial \varphi} = \frac{\partial Q^\mu}{\partial \varphi} \frac{\partial}{\partial Q^\mu} + \frac{\partial P_\nu}{\partial \varphi} \frac{\partial}{\partial P_\nu} = \frac{\partial}{\partial Q^2},
\end{eqnarray}
\end{subequations}
where we have used the fact that
\begin{equation}
\frac{\partial Q^1}{\partial t} = -1, \quad \frac{\partial Q^0}{\partial t} = \frac{\partial Q^2}{\partial t} = 0, \quad  \frac{\partial Q^2}{\partial \varphi} = 1, \quad  \frac{\partial Q^0}{\partial \varphi} = \frac{\partial Q^1}{\partial \varphi} = 0.
\end{equation}
Thus, a distribution function corresponding to a stationary and axially symmetric configuration can be sought in the form
\begin{equation}
f = f(m,E,l_z).
\end{equation}

\subsection{Volume element in the momentum space}

We compute the components of $J_\mu(t,r,\varphi)$ and $T_{\mu\nu}(t,r,\varphi)$ using Eqs.\ (\ref{jmutmunusurface}). Since the distribution function can be expressed in terms of the constants of motion $m$, $E$, and $l_z$ (or $m$, $E$, and $l$), it is convenient to introduce a suitable change of the momentum (integration) variables. Consider a transformation $(p_t,p_r,p_\varphi) \mapsto (m,E,l)$. For the motion restricted to the equatorial plane, we get in explicit terms,
\begin{eqnarray}
    m^2 & = & p_t^2 - p_r^2 - \frac{(a p_r + p_\varphi)^2}{r^2} + \frac{2 M (p_r - p_t)^2}{r}, \\
    E & = & - p_t, \\
    l & = & \epsilon_\sigma (p_\varphi + a p_t).
\end{eqnarray}
This gives
\begin{equation}
\frac{\partial (m^2,E,l)}{\partial (p_t,p_r,p_\varphi)} = - \frac{2 \epsilon_\sigma (\Delta p_r + 2 M r p_t + a p_\varphi)}{r^2} = - \frac{2 \epsilon_\sigma \epsilon_r \sqrt{R(r)}}{r^2}.
\end{equation}
Thus
\begin{equation}
dm^2 dE dl = \frac{2 \sqrt{R(r)}}{r^2} dp_t dp_r dp_\varphi,
\end{equation}
and the volume element $\sqrt{- \mathrm{det} \, \gamma^{\mu \nu}} dp_t dp_r dp_\varphi$ reads
\begin{equation}
\sqrt{- \mathrm{det} \, \gamma^{\mu \nu}} dp_t dp_r dp_\varphi =  \frac{r}{\sqrt{R(r)}} m dm dE dl.
\end{equation}

\subsection{Dimensionless coordinates}

In the following we will use dimensionless variables $\xi$, $\alpha$, $\varepsilon$, $\lambda$, and $\lambda_z$ defined by
\begin{equation} r = M \xi, \quad a = M \alpha, \quad E = m \varepsilon, \quad l = M m \lambda, \quad l_z = M m \lambda_z.
\end{equation}
Using the above definitions, one has
\begin{equation}
R(r) = M^4 m^2 \tilde R(\xi), \quad \Delta = M^2 \tilde \Delta,
\end{equation}
where
\begin{equation}
\tilde R(\xi) = \left[ \varepsilon (\xi^2 + \alpha^2) - \alpha \lambda_z \right]^2 - \tilde \Delta (\lambda^2 + \xi^2)
\end{equation}
and
\begin{equation}
\tilde \Delta = \xi^2 - 2 \xi + \alpha^2.
\end{equation}
At the equatorial plane $\lambda_z = \epsilon_\sigma \lambda + \alpha \varepsilon$, and thus
\begin{equation}
\tilde R(\xi) = \left( \varepsilon \xi^2 - \epsilon_\sigma \alpha \lambda \right)^2 - \tilde \Delta (\lambda^2 + \xi^2).
\end{equation}
In the same fashion
\begin{equation}
\sqrt{- \mathrm{det} \, \gamma^{\mu \nu}} dp_t dp_r dp_\varphi =  \frac{\xi m^2}{\sqrt{\tilde R(\xi)}} dm d\varepsilon d\lambda.
\end{equation}
The momenta can be expressed as
\begin{subequations}
\begin{eqnarray}
    p_t & = & - m \varepsilon, \\
    p_r & = & m \frac{2 \xi \varepsilon - \alpha \lambda_z + \epsilon_r \sqrt{\tilde R(\xi)}}{\tilde \Delta} = m \frac{2 \xi \varepsilon - \epsilon_\sigma \alpha \lambda - \alpha^2 \varepsilon + \epsilon_r \sqrt{\tilde R(\xi)}}{\tilde \Delta}, \\
    p_\varphi & = & M m (\epsilon_\sigma \lambda + \alpha \varepsilon).
\end{eqnarray}
\end{subequations}
Note, however, that
\begin{equation}
p^r = \epsilon_r m \frac{\sqrt{\tilde R(\xi)}}{\xi^2},
\end{equation}
which has the same form, as in the Boyer-Lindquist coordinates.

In the remainder of this paper we will also use dimensionless expressions for the horizon radii $\xi_\pm = 1 \pm \sqrt{1 - \alpha^2}$.

To normalize our solutions, we will also refer to the following expression for $- J_\mu J^\mu$ in the case of matter distribution confined to the equatorial plane:
\begin{eqnarray}
    - J_\mu J^\mu & = & \frac{1}{r \Delta} \left\{ a^2 (2M + r) J_t^2 + r^3 [J_t^2 - (J^r)^2] - (r - 2M) J_\varphi^2 + 4 M a J_t J_\varphi \right\} \nonumber \\
    & = & \frac{1}{\xi \tilde \Delta} \left\{ \alpha^2  (2 + \xi) J_t^2 + \xi^3 [J_t^2 - (J^r)^2] - (\xi - 2) \frac{J_\varphi^2}{M^2} + 4 \alpha J_t \frac{J_\varphi}{M} \right\}.
    \label{JJ}
\end{eqnarray}
We define the (covariant) particle number surface density as $n_s = \sqrt{-J_\mu J^\mu}$, and the surface mass density as $\rho_s = m n_s = m \sqrt{-J_\mu J^\mu}$.

\subsection{Radial motion}

\subsubsection{Radial potential and its critical points}

The types of orbits confined to the equatorial plane are governed by the radial potential $\tilde R(\xi)$. We will now summarize its relevant properties, basing on the analysis given in \cite{CMO2022,RO2023}. It is known (see, e.g., \cite{RO2023}) that the radial potential can be conveniently factorized as
\begin{equation}
\tilde R(\xi) = \xi^4 \left[ \varepsilon - W_-(\xi) \right] \left[ \varepsilon - W_+(\xi) \right],
\end{equation}
where
\begin{equation}
W_\pm(\xi) = \frac{\epsilon_\sigma \alpha \lambda}{\xi^2} \pm \frac{\sqrt{\tilde \Delta (\xi^2 + \lambda^2)}}{\xi^2}.
\end{equation}
The motion is only allowed in a region where $\tilde R(\xi) \ge 0$. Outside the black hole horizon, i.e., for $\xi \ge \xi_+$, we have $W_-(\xi) \le W_+(\xi)$, with the equality occurring for $\xi = \xi_+$. Hence, for $\xi \ge \xi_+$ the condition $\tilde R(\xi) \ge 0$ implies that either $\varepsilon \ge W_+$ or $\varepsilon \le W_-$. It can be shown that only the former condition allows for future directed timelike geodesics \cite{RO2023}. Hence, it is sufficient to analyse the condition $\varepsilon \ge W_+(\xi)$. For $\alpha = 0$, the expression for $W_+(\xi)$ reduces to the well-known effective radial potential corresponding to the (timelike) geodesic motion in the Schwarzschild spacetime, i.e.,
\begin{equation}
\left. W_+(\xi) \right|_{\alpha = 0} =\sqrt{1 - \frac{2}{\xi}} \sqrt{1 + \frac{\lambda^2}{\xi^2}}.
\end{equation}

Let us start the analysis of $W_+$ with a discussion of its critical points. Suppose that $\xi > \xi_+ = 1+\sqrt{1-\alpha^2}$ is a critical point of $W_+$, i.e., 
\begin{eqnarray}
    \frac{d W_+}{d\xi}= \frac{-2 \alpha \epsilon_{\sigma} \lambda \sqrt{(\xi^2 - 2\xi + \alpha^2)(\xi^2+\lambda^2)} + \xi^2 (\xi-\alpha^2)-(\xi^2-3\xi+2\alpha^2)\lambda^2}{\xi^3 \sqrt{(\xi^2 - 2\xi + \alpha^2)(\xi^2+\lambda^2)}}=0.
\end{eqnarray}
This is, of course, equivalent to the condition
\begin{equation}
\label{equationlambda}
\xi^2 (\xi-\alpha^2)-(\xi^2-3\xi+2\alpha^2)\lambda^2 = 2 \alpha \epsilon_{\sigma} \lambda \sqrt{(\xi^2 - 2\xi + \alpha^2)(\xi^2+\lambda^2)}.
\end{equation}
By computing the square of both sides we get the following bi-quadratic equation for $\lambda$:
\begin{eqnarray}\label{bi-quadratic}
    \mathcal{A} \lambda^4 - 2 \mathcal{B} \lambda^2 + \mathcal{C} = 0,
\end{eqnarray}
where
\begin{subequations}
\begin{eqnarray}
\mathcal{A} & = & \xi^2(\xi-3)^2 - 4 \alpha^2 \xi,  \\
\mathcal{B} & = & \xi^4(\xi-3)+\alpha^2 \xi^3(\xi+1),  \\
\mathcal{C} & = & \xi^4(\xi - \alpha^2)^2.
\end{eqnarray}
\end{subequations}
The discriminant of Eq.\ (\ref{bi-quadratic}) reads
\begin{eqnarray}
    D := \mathcal{B}^2 -\mathcal{A} \mathcal{C} = 4 \alpha^2 \xi^5 \tilde \Delta^2,
\end{eqnarray}
and it is positive for $\xi > \xi_+$. Thus, Eq.\ (\ref{bi-quadratic}) has two reals roots, given by $\lambda^2=(\mathcal{B} \pm \sqrt{D})/\mathcal{A}$. Explicit formulas for $\lambda^2$ can be simplified using a factorization of $\mathcal{A}$ in the form
\begin{equation}
\mathcal{A} = \left[ \xi(\xi-3) + 2 \alpha \sqrt{\xi} \right] \left[ \xi(\xi-3) - 2 \alpha \sqrt{\xi} \right].
\end{equation}
This yields
\begin{eqnarray}\label{lambda_square}
    \lambda^2=\frac{\xi^2 (\sqrt{\xi} - \epsilon \alpha)^2}{\xi(\xi-3) + 2 \epsilon \alpha \sqrt{\xi}},
\end{eqnarray}
where $\epsilon = \pm 1$ is a sign. A real solution for $\lambda$ can only exist, if $\xi(\xi-3) + 2  \epsilon \alpha \sqrt{\xi}>0$. This gives two possible solutions for $\lambda$:
\begin{equation}
    \lambda = \frac{\xi (\sqrt{\xi} - \epsilon \alpha)}{\sqrt{\xi(\xi-3) + 2 \epsilon \alpha \sqrt{\xi}}}, \quad \lambda = - \frac{\xi (\sqrt{\xi} - \epsilon \alpha)}{\sqrt{\xi(\xi-3) + 2 \epsilon \alpha \sqrt{\xi}}}.
\end{equation}
By a straightforward calculation one can check that both solutions satisfy Eq.\ (\ref{equationlambda}), provided that $\epsilon = \epsilon_\sigma$ in the first one, and $\epsilon = - \epsilon_\sigma$ in the second case. On the other hand, for $\xi > \xi_+ = 1 + \sqrt{1- \alpha^2}$, one has $\sqrt{\xi} - |\alpha| > 0$ (for $-1 \le \alpha \le 1$). Thus the second of the solutions for $\lambda$ cannot be positive. In summary, the expression for $\lambda$ is given as
\begin{eqnarray}
\label{lambdacritical}
    \lambda=\frac{\xi(\sqrt{\xi} - \epsilon_{\sigma} \alpha)}{\sqrt{h (\xi)}},
\end{eqnarray}
where $h$ is a smooth function $h \colon (\xi_+,\infty) \rightarrow \mathbb R$ defined by
\begin{eqnarray}
    h (\xi) := \xi(\xi-3) + 2 \epsilon_\sigma \alpha \sqrt{\xi}.
\end{eqnarray}

Substituting the value of $\lambda$ given by Eq.\ (\ref{lambdacritical}) into the expression for $W_+$, we get the value
\begin{equation}
    W_+ = \frac{\xi^2 - 2\xi + \epsilon_\sigma \alpha  \sqrt{\xi}}{\xi \sqrt{h (\xi)}},
\end{equation}
derived under the assumptions that $\xi \ge \xi_+$ and $h(\xi) > 0$. This yields explicit expressions for the values of $\lambda$ and the energy $\varepsilon$ associated with critical circular orbits, which we denote as
\begin{equation}
    \lambda_c(\xi) := \frac{\xi(\sqrt{\xi} - \epsilon_{\sigma} \alpha)}{\sqrt{h (\xi)}}, \quad \varepsilon_c(\xi) := \frac{\xi^2 - 2\xi + \epsilon_\sigma \alpha  \sqrt{\xi}}{\xi \sqrt{h (\xi)}}.
\end{equation}

\subsubsection{Behavior of \texorpdfstring{$\lambda_c(\xi)$}{lambda c} and \texorpdfstring{$\varepsilon_c(\xi)$}{epsilon c}}

\begin{figure}[t]
\subfigure[]{\label{fig:2a}\includegraphics[width=88mm]{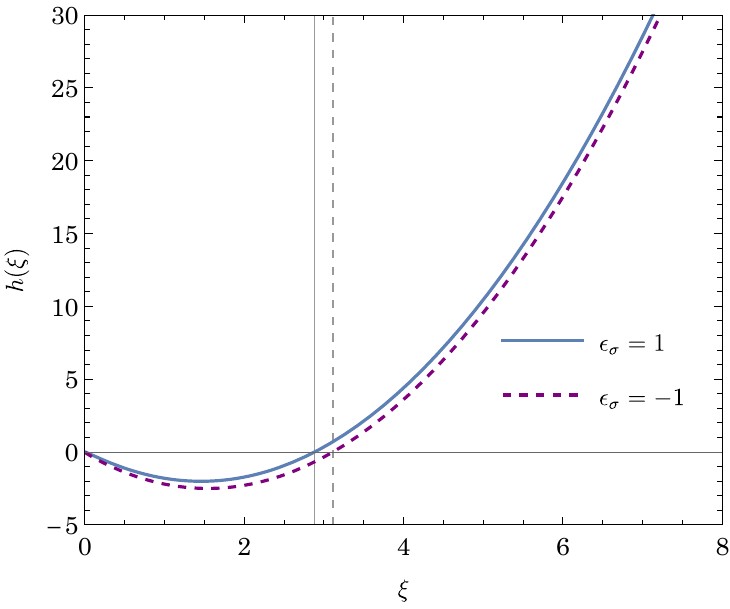}} 
\subfigure[]{\label{fig:2b}\includegraphics[width=88mm]{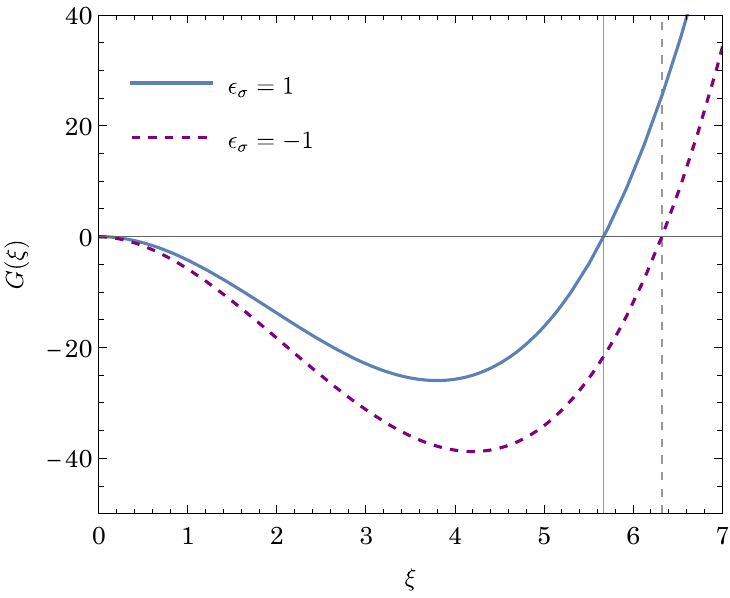}}
\caption{Sample plots of $h(\xi)$ (a) and $G(\xi)$ (b) for $\alpha=1/10$. Locations of circular photon orbits $\xi_\mathrm{ph}$(left panel) and marginally stable orbits $\xi_\mathrm{ms}$(right panel) are  marked with vertical lines.}
\label{lambdasph}
\end{figure}

Rioseco and Sarbach have shown in \cite{RO2023} (Lemma 19) that $h(\xi)$ has only one zero outside the horizon, i.e., for $\xi > \xi_+$. It is given by
\begin{equation}
\xi_\mathrm{ph} =  2 + 2 \cos \left[ \frac{2}{3} \arccos(-\epsilon_\sigma \alpha) \right],
\end{equation}
and it is equal to the dimensionless radius of a circular photon orbit within the equatorial plane. The function $h$ is negative for $\xi \in (\xi_+,\xi_\mathrm{ph})$ and positive for $\xi > \xi_\mathrm{ph}$. That means that both $\lambda_c$ and $\varepsilon_c$ diverge at $\xi = \xi_\mathrm{ph}$ and that they are defined (as real-valued functions) only for $\xi > \xi_\mathrm{ph}$. Sample graphs of $h(\xi)$ are shown in Fig.\ \ref{lambdasph}.

Both $\lambda_c$ and $\varepsilon_c$ are positive for $\xi > \xi_\mathrm{ph}$. It is easy to see that $\lim_{\xi \to \xi_\mathrm{ph}^+} \lambda_c = + \infty$. One can also show that $\lambda_c$ attains a minimum at $\xi = \xi_\mathrm{ms} > \xi_\mathrm{ph}$. To determine this location, we compute
\begin{eqnarray}
    \frac{d \lambda_c}{d\xi} = \frac{G (\xi)}{2 h (\xi)^{3/2}\sqrt{\xi}}, \quad G(\xi)=\xi \tilde \Delta(\xi)- [h (\xi)- \xi(\xi-1)]^2.
\end{eqnarray}
The function $G(\xi)$ has a zero within the interval $(\xi_\mathrm{ph},\infty)$, given by \cite{Bardeen1972}
\begin{equation}
\label{xims}
    \xi_\mathrm{ms} = 3 + Z_2 - \epsilon_\sigma \alpha \sqrt{\frac{(3 - Z_1)(3 + Z_1 + 2 Z_2)}{\alpha^2}},
\end{equation}
where
\begin{subequations}
\begin{eqnarray}
    Z_1 & = & 1 + (1 - \alpha^2)^{1/3} \left[ (1 + \epsilon_\sigma \alpha)^{1/3} + (1 - \epsilon_\sigma \alpha)^{1/3} \right], \\
    Z_2 & = & \sqrt{3 \alpha^2 + Z_1^2}.
\end{eqnarray}
\end{subequations}
It can also be proved (\cite{RO2023}, Lemma 20) that $G(\xi)$ is negative for $\xi < \xi_\mathrm{ms}$ and positive for $\xi > \xi_\mathrm{ms}$ (see also Fig.\ \ref{fig:2b}). It follows that as $\xi$ increases from $\xi_\mathrm{ph}$ to $+\infty$, the value of $\lambda_c$ decreases from $+\infty$ to a minimum value at $\xi = \xi_\mathrm{ms}$, and then increases monotonically to $+\infty$.

Some of the properties of $\lambda_c$ are reflected in the behavior of $\varepsilon_c$. Both $\lambda_c$ and $\varepsilon_c$ diverge at $\xi=\xi_\mathrm{ph}$. In addition, a simple calculation gives
\begin{eqnarray}
    \frac{d \varepsilon_c}{d \xi} = \xi^{-\frac{3}{2}} \frac{d \lambda_c}{d\xi},
\end{eqnarray}
which shows that $\varepsilon_c(\xi)$ also attains a minimum at $\xi = \xi_\mathrm{ms}$. On the other hand, while $\lambda_c$ tends to $+\infty$, as $\xi \to \infty$, we have $\lim_{\xi \to \infty} \varepsilon_c = +1$. Sample graphs of $\lambda_c(\xi)$ and $\varepsilon_c(\xi)$ are shown in Fig.\ \ref{lambdasph,epsilonsph}.

\begin{figure}[t]
\subfigure[]{\label{fig:3a}\includegraphics[width=88mm]{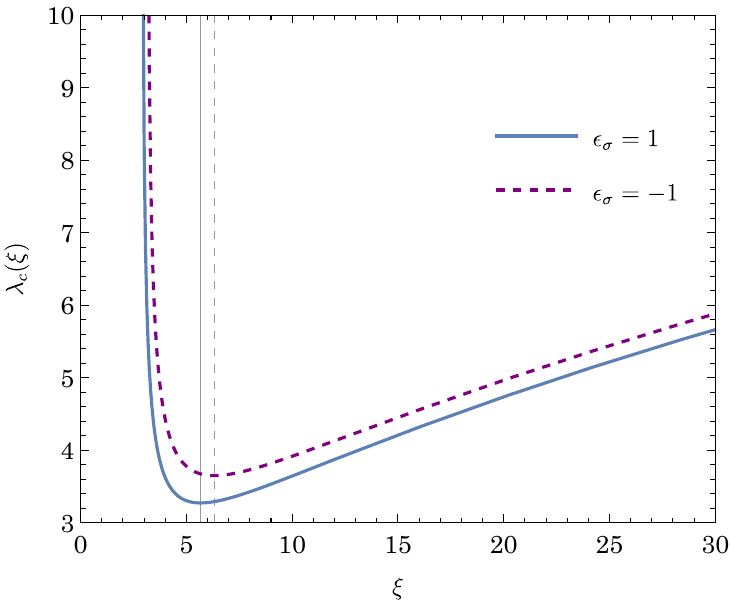}} 
\subfigure[]{\label{fig:3b}\includegraphics[width=88mm]{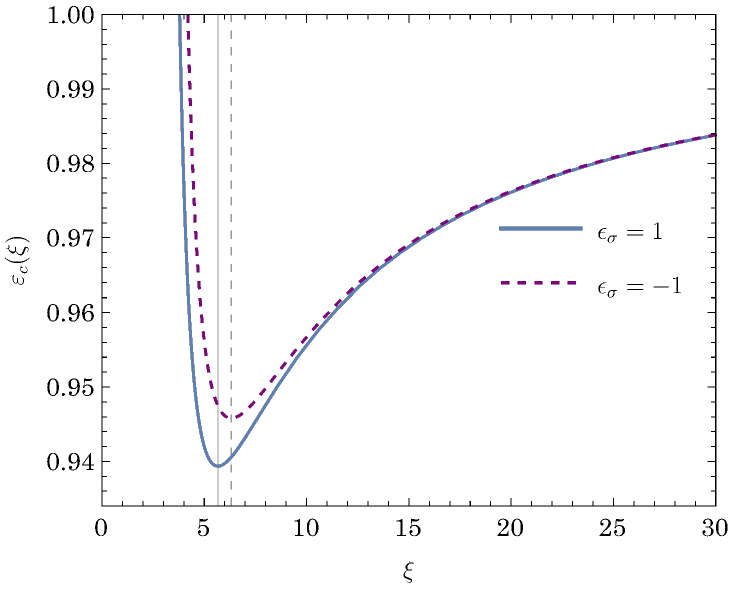}}
\caption{Sample plots of $\lambda_{c}(\xi)$ (a) and $\varepsilon_{c}(\xi)$ (b) for   $\alpha=1/10$. Locations of marginally stable orbits $\xi_\mathrm{ms}$ are marked with vertical lines.}
\label{lambdasph,epsilonsph}
\end{figure}

\subsubsection{Global behavior of the radial potential}

We see that for any $\lambda > \lambda_c(\xi_\mathrm{ms})$, there exist precisely two radii $\xi_\mathrm{max}$ and $\xi_\mathrm{min}$ such that $\xi_\mathrm{ph} < \xi_\mathrm{max} < \xi_\mathrm{ms} < \xi_\mathrm{min}$ and $\lambda_c(\xi_\mathrm{max}) = \lambda_c(\xi_\mathrm{min}) = \lambda$.
A lengthy but simple calculation shows that
\begin{eqnarray}
    \left. \frac{d^2 W_+}{d \xi^2} \right|_{\lambda = \lambda_c} = \frac{G(\xi)}{\xi^3 \tilde \Delta (\xi) \sqrt{h (\xi)}},
\end{eqnarray}
provided that $h(\xi) > 0$. Thus, $W_+$ has a local maximum at $\xi_\mathrm{max}$ and a local minimum at $\xi_\mathrm{min}$.

Our reasoning can now be summarized as follows.
\begin{itemize}
    \item  If $0 < \lambda < \lambda_c(\xi_\mathrm{ms})$, $W_{+}(\xi)$ is a monotonically increasing function of $\xi$.
    \item  If $\lambda > \lambda_c(\xi_\mathrm{ms})$, $W_{+}(\xi)$ has a local maximum at $\xi_\mathrm{max}$ $\in$ $(\xi_\mathrm{ph},\xi_\mathrm{ms})$ and a local minimum at $\xi_\mathrm{min}$ $\in$ $(\xi_\mathrm{ms},+\infty)$. The corresponding energy values are $\varepsilon_\mathrm{max}:=\varepsilon_c(\xi_\mathrm{max})$ and $\varepsilon_\mathrm{min}:=\varepsilon_c(\xi_\mathrm{min})$. As $\lambda$ increases from $\lambda_c(\xi_\mathrm{ms})$ to $+\infty$, $\xi_\mathrm{max}$ decreases from $\xi_\mathrm{ms}$ to $\xi_\mathrm{ph}$, $\varepsilon_\mathrm{max}$ increases from $\varepsilon_\mathrm{ms} := \varepsilon_c(\xi_\mathrm{ms})$ to $+\infty$, and $\varepsilon_\mathrm{min}$ increases from $\varepsilon_\mathrm{ms}$ to 1.  
\end{itemize}

\subsection{Phase-space ranges}

We will now discuss the ranges of phase-space parameters, relevant for stationary equatorial accretion occurring from a finite radius $\xi_0$. We take into account all generic equatorial timelike geodesic trajectories that can reach $\xi = \xi_0$. Consider a generic test particle moving from $\xi = \xi_0$ inward (i.e., toward the black hole). This particle can either plunge into the black hole (this means that no radial turning points occur between $\xi_+$ and $\xi_0$) or, if its angular momentum is sufficiently high, it can be scattered by the centrifugal potential and reach $\xi = \xi_0$ again. We will refer to those two possibilities as corresponding to absorbed and scattered trajectories, respectively. In both cases, the particle can move along a bound orbit, that is, a trajectory characterized by an energy $\varepsilon < 1$. Such an orbit can never reach $\xi = +\infty$. This makes a key technical difference compared to our previous study \cite{CMO2022}, in which we investigated the accretion occurring from $\xi = +\infty$. An analogous difference between the two models (with boundary conditions at a finite radius and at infinity) was analyzed in \cite{olivierfinite} for the Schwarzschild geometry. Our restriction to ``generic'' trajectories means that we exclude a zero-measure set of orbits that inspiral toward critical circular trajectories (see, e.g., \cite{OShaughnessy2003,Levin2009}).

\subsubsection{Minimum energy at a given radius}

As the first step of our analysis, we answer the following question. Fix the radius $\xi$. What is the minimum allowed energy $\varepsilon$ for a particle that can reach $\xi$? Equivalently---what is the minimum possible value of $W_+$ at $\xi$? For $\alpha = 0$, $W_+$ is a growing function of $\lambda \in [0,+\infty)$, and the minimum value is obtained for $\lambda = 0$. This is no longer true for $\alpha \neq 0$. Note that
\begin{equation}
\label{dWpdl}
    \frac{dW_+}{d\lambda} = \frac{\epsilon_\sigma \alpha \sqrt{\lambda^2 + \xi^2} + \lambda \sqrt{\tilde \Delta}}{\xi^2 \sqrt{\lambda^2 + \xi^2}}.
\end{equation}
A minimum of $W_+$ with respect to $\lambda$ can occur for $- \epsilon_\sigma \alpha \sqrt{\lambda^2 + \xi^2} = \lambda \sqrt{\tilde \Delta}$. This can happen for a positive $\lambda$ only if $\epsilon_\sigma \alpha$ remains negative and $\xi > 2$. The solution to the equation $dW_+/d\lambda = 0$ is then given by
\begin{equation}
\label{lambdamin}
    \lambda = - \frac{\epsilon_\sigma \alpha \sqrt{\xi}}{\sqrt{\xi - 2}}.
\end{equation}

If $\epsilon_\sigma \alpha \ge 0$, Eq.\ (\ref{dWpdl}) gives $dW_+/d\lambda \ge 0$, so that $W_+$ is a non-decreasing function of $\lambda$. In this case, the minimum of $W_+$ in the range of $\lambda \in [0,+\infty)$ is attained for $\lambda = 0$, and the minimum energy at a given radius $\xi$ reads
\begin{equation}
W_\mathrm{min} = \left. W_+(\xi) \right|_{\lambda = 0} = \frac{\sqrt{\tilde \Delta}}{\xi} = \sqrt{1 - \frac{2}{\xi} + \frac{\alpha^2}{\xi^2}}.
\end{equation}
If $\epsilon_\sigma \alpha < 0$ and $\xi > 2$, the minimum is attained for $\lambda$ given by Eq.\ (\ref{lambdamin}). The lower bound on the allowed energy is then given by
\begin{equation}
    W_\mathrm{min} = \left. W_+(\xi) \right|_{\lambda = -\epsilon_\sigma \alpha \sqrt{\xi}/\sqrt{\xi - 2}} = \sqrt{1 - \frac{2}{\xi}}.
\end{equation}
If $\epsilon_\sigma \alpha < 0$, but $\xi_+ < \xi < 2$ (i.e., $\xi$ refers to a point inside the ergosphere), a straightforward calculation shows that
\begin{equation}
    - |\alpha| \sqrt{\lambda^2 + \xi^2} + \lambda \sqrt{\xi^2 - 2 \xi + \alpha^2} < 0.
\end{equation}
Thus, $W_+$ is a decreasing function of $\lambda \in [0,+\infty)$. As it is well known, in this case $W_+$ can become negative \cite{Contopoulos1984,Grib2014,Vertogradov2015}, allowing for orbits with negative energies. Note that for a fixed $\xi \neq 0$, the limit $\lim_{\lambda \to +\infty} W_+$ is $\pm \infty$, depending on the sign of $\epsilon_\sigma \alpha + \sqrt{\tilde \Delta}$. For $\xi = 2$, one has $\lim_{\lambda \to +\infty} W_+ = 0$.

In summary, for $\xi > 2$ we have
\begin{eqnarray}
    W_\mathrm{min}(\xi,\epsilon_\sigma) = \begin{cases}
        \sqrt{1 - \frac{2}{\xi} + \frac{\alpha^2}{\xi^2}}, & \epsilon_\sigma \alpha \ge 0, \\
        \sqrt{1 - \frac{2}{\xi}}, & \epsilon_\sigma \alpha < 0.
    \end{cases}
\end{eqnarray}
For $\alpha = 0$ one obtains simply $W_\mathrm{min} = \sqrt{1 - 2/\xi}$.

\subsubsection{Critical energy, bounds on \texorpdfstring{$\lambda$}{lambda}}

Let us return to the characterization of phase-space parameters describing the equatorial motion of particles in our disk models. The condition that the disk particles can reach an outer edge located at the radius $\xi_0$ fixes the minimal allowed energy $\varepsilon = W_\mathrm{min}(\xi_0,\epsilon_\sigma)$.

\begin{figure}[t]
\subfigure[]{
\includegraphics[width=.49\textwidth]{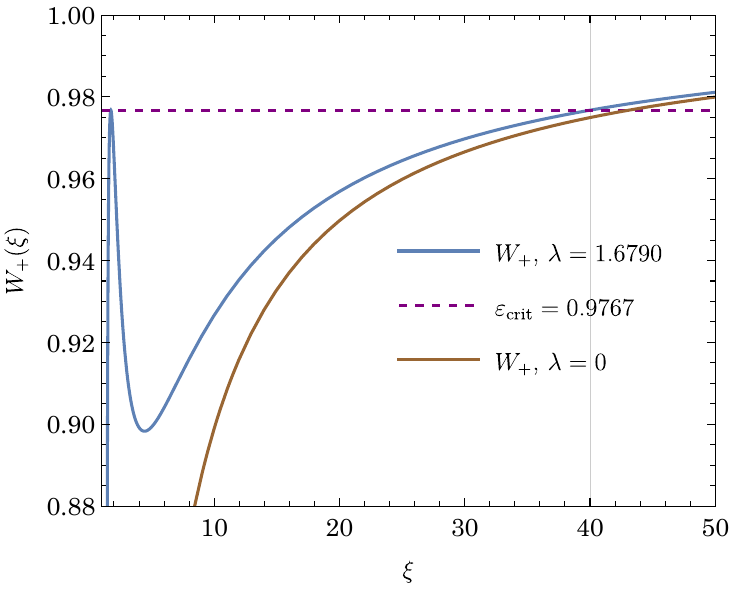}}
\subfigure[]{
\includegraphics[width=.49\textwidth]{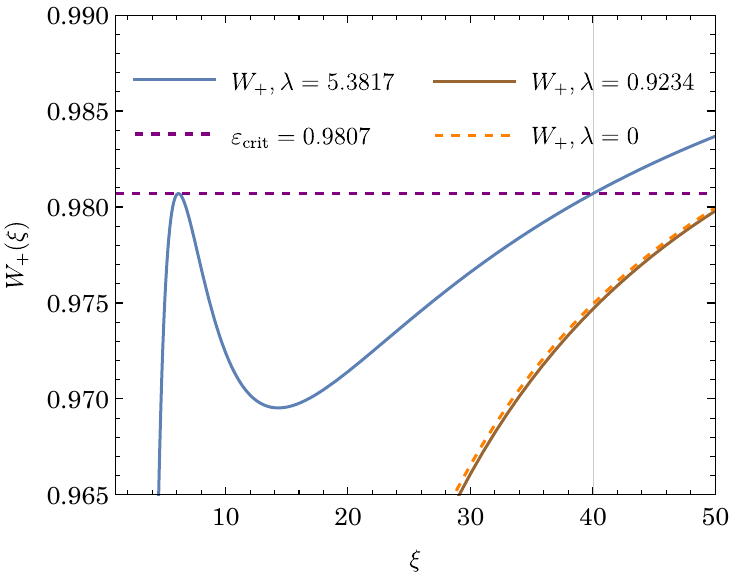}}
\caption{\label{criticalenergy}An illustration of the critical energy $\varepsilon_\mathrm{crit}$. In this example $\alpha = 9/10$, $\epsilon_\sigma = +1$ (a), $\epsilon_\sigma = -1$ (b), the outer radius is set to $\xi_0 = 40$.}
\end{figure}

We will further distinguish between particles absorbed by the black hole and those scattered by the centrifugal barrier. The distinction between these two classes can be made basing on a critical energy $\varepsilon_\mathrm{crit}(\xi_0,\epsilon_\sigma)$, which we define as follows. Consider possible values of $W_+$ at $\xi = \xi_0$. Let $\lambda_\mathrm{min}$ correspond to the smallest allowed value $W_+(\xi_0) = W_\mathrm{min}(\xi_0,\epsilon_\sigma)$ (be it $\lambda_\mathrm{min} = 0$ for $\epsilon_\sigma \alpha \ge 0$ or $\lambda_\mathrm{min} = - \epsilon_\sigma \alpha \sqrt{\xi}/\sqrt{\xi - 2}$ for $\epsilon_\sigma \alpha < 0$). Suppose that we now increase the value of $\lambda$, starting from $\lambda = \lambda_\mathrm{min}$. Clearly, the value $W_+(\xi_0)$ also increases. As long as $\lambda < \lambda_c(\xi_\mathrm{ms})$, $W_+(\xi)$ remains a monotonic (growing) function of $\xi$. Thus the whole region of $\xi_+ \le \xi \le \xi_0$ is available for the motion, and the particle starting at $\xi = \xi_0$ plunges into the black hole. For $\lambda > \lambda_c(\xi_\mathrm{ms})$, the potential $W_+(\xi)$ has a local maximum; its value at this maximum, denoted as $\varepsilon_\mathrm{max}$, grows from $\varepsilon_c(\xi_\mathrm{ms})$ to infinity, as $\lambda$ increases. There is a critical value of $\lambda$, denoted as $\lambda_\mathrm{crit}(\xi_0,\epsilon_\sigma)$, for which the value of the potential $W_+$ at the local maximum, i.e., $\varepsilon_\mathrm{max}$, equals to $W_+(\xi_0)$. This value depends obviously on the choice of $\xi_0$. The corresponding critical energy $\varepsilon = W_+(\xi_0) = \varepsilon_\mathrm{max}$ will be denoted by $\varepsilon_\mathrm{crit}(\xi_0,\epsilon_\sigma)$. This situation is illustrated in Fig.\ \ref{criticalenergy}. Increasing $\lambda$ even further leads to a case in which $\varepsilon_\mathrm{max}$ exceeds $W_+(\xi_0)$. In this situation, both scattered and absorbed trajectories are allowed. Trajectories with $W_+(\xi_0) \le \varepsilon < \varepsilon_\mathrm{max}$ are scattered; those characterized by $\varepsilon > \varepsilon_\mathrm{max}$ are absorbed by the black hole.

On the other hand, $\lambda$ cannot grow indefinitely. The upper limit on the allowed value of $\lambda$ for a particle of energy $\varepsilon$ at radius $\xi$ follows from the requirement that $W_+(\xi) \le \varepsilon$. It reads, for $\xi > 2$,
\begin{equation}
    \lambda_\mathrm{max}  (\xi,\varepsilon,\epsilon_\sigma) =\frac{\xi}{\xi-2}  \Bigg \{ \sqrt{\alpha^2 \left(\varepsilon^2 + \frac{2}{\xi}-1 \right) +(\xi-2)\left[ \xi(\varepsilon^2-1) + 2 \right] }-\epsilon_\sigma \alpha \varepsilon \Bigg \}
    \label{lambdamax}
\end{equation}
(see a discussion below). The critical energy $\varepsilon_\mathrm{crit}$ can also be characterized in terms of $\lambda_\mathrm{max}(\xi,\varepsilon,\epsilon_\sigma)$, by a condition that $\lambda_\mathrm{max}(\xi_0,\varepsilon_\mathrm{crit}(\xi_0,\epsilon_\sigma),\epsilon_\sigma) = \lambda_c(\varepsilon_\mathrm{crit}(\xi_0,\epsilon_\sigma))$. An analytic formula for $\varepsilon_\mathrm{crit}(\xi_0,\epsilon_\sigma)$ can be found for $\alpha = 0$. For $\alpha \neq 0$, we compute $\varepsilon_\mathrm{crit}(\xi_0,\epsilon_\sigma)$ numerically.

There is a subtlety related to the minimum allowed value of $\lambda$, which becomes important for $\epsilon_\sigma \alpha < 0$. The radial potential $\tilde R(\xi)$ can be expressed as the following quadratic function of $\lambda$:
\begin{equation}
    \tilde R = - \xi(\xi-2) \lambda^2 - 2 \epsilon_\sigma \alpha \varepsilon \xi^2 \lambda + \xi^2 (\varepsilon^2 \xi^2 - \tilde \Delta).
\end{equation}
Since the leading coefficient is simply $-\xi(\xi - 2)$, for $\xi > 2$ the condition $\tilde R > 0$ can only be satisfied provided that $\tilde R$ has two real zeros with respect to $\lambda$. In this case $\tilde R$ can be factored as
\begin{equation}
    \tilde R = - \xi (\xi - 2) (\lambda - \lambda_\mathrm{min})(\lambda - \lambda_\mathrm{max}),
\end{equation}
where $\lambda_\mathrm{max}$ is given by Eq.\ (\ref{lambdamax}) and $\lambda_\mathrm{min}$ reads
\begin{equation}
     \lambda_\mathrm{min}  (\xi,\varepsilon,\epsilon_\sigma) = \frac{\xi}{\xi-2}  \Bigg \{ - \sqrt{\alpha^2 \left(\varepsilon^2 + \frac{2}{\xi}-1 \right) +(\xi-2)\left[ \xi(\varepsilon^2-1) + 2 \right] } - \epsilon_\sigma \alpha \varepsilon \Bigg \}.
\end{equation}
The condition $\tilde R(\xi) \ge 0$ now yields $\lambda_\mathrm{min} \le \lambda \le \lambda_\mathrm{max}$. It is easy to show that for $\epsilon_\sigma \alpha > 0$ and $\xi > 0$, one has $\lambda_\mathrm{min} < 0$. Given our convention that $\lambda \ge 0$, we get the allowed range $0 \le \lambda \le \lambda_\mathrm{max}$. The situation can be different for $\epsilon_\sigma \alpha < 0$. In this case, $\lambda_\mathrm{min} \ge 0$ for $\varepsilon \le \sqrt{\tilde \Delta}/\xi$, and $\lambda_\mathrm{min} < 0$ for $\varepsilon > \sqrt{\tilde \Delta}/\xi$. That means, in turn, that for $\epsilon_\sigma \alpha < 0$ and $\varepsilon \le \sqrt{\tilde \Delta}/\xi$ the relevant allowed range of $\lambda$ is bounded by $\lambda_\mathrm{min} \le \lambda \le \lambda_\mathrm{max}$. This reasoning implies the following definition:
\begin{equation}
\Lambda_\mathrm{min}(\xi,\varepsilon,\epsilon_\sigma) :=
    \begin{cases}
        \lambda_\mathrm{min}(\xi,\varepsilon,\epsilon_\sigma)   &   \epsilon_{\sigma} \alpha  \leq 0, \, \varepsilon \leq \frac{\sqrt{\tilde \Delta}}{\xi}, \\
        0 & \text{otherwise}.
    \end{cases}
\end{equation}

The case with $\xi < 2$ is much less important, as we always put the outer edge $\xi_0 > 2$.

\subsubsection{Phase-space parameters for accretion occurring from a finite radius}

The discussion from the previous paragraph yields the following characterization of the relevant phase-space parameters.
\begin{itemize}
    \item Absorbed orbits:
    \begin{equation}
    \begin{split}
        W_\mathrm{min}(\xi_0,\epsilon_\sigma) \le \varepsilon \le \varepsilon_\mathrm{crit}(\xi_0,\epsilon_\sigma) \quad \text{and} \quad \Lambda_\mathrm{min}(\xi_0,\varepsilon,\epsilon_\sigma) \le \lambda \le \lambda_\mathrm{max}(\xi_0,\varepsilon,\epsilon_\sigma),
        \end{split}
    \end{equation}
    or
    \begin{equation}
    \varepsilon_\mathrm{crit}(\xi_0,\epsilon_\sigma) < \varepsilon < +\infty \quad \text{and} \quad 0 \le \lambda < \lambda_c(\varepsilon,\epsilon_{\sigma}).
    \end{equation}
    \item Scattered orbits:
    \begin{equation}
    \label{scattcondition}
    \varepsilon_\mathrm{min}(\xi,\epsilon_\sigma) < \varepsilon < +\infty \quad \text{and} \quad \lambda_c(\varepsilon,\epsilon_{\sigma}) < \lambda \le \Lambda_\mathrm{max}(\xi,\xi_0,\varepsilon,\epsilon_\sigma).
    \end{equation}
\end{itemize}
In the above expressions, we have defined the minimum energy of a particle moving along a scattered orbit. It reads:
\begin{eqnarray}
\varepsilon_\mathrm{min}  (\xi,\epsilon_\sigma) =
    \begin{cases}
        \infty &      \xi \le \xi_\mathrm{ph}, \\
        \varepsilon_c(\xi,\epsilon_{\sigma})   &   \xi_\mathrm{ph} < \xi < \xi_1, \\
        \varepsilon_\mathrm{crit}(\xi_0,\epsilon_\sigma) &       \xi \ge \xi_1.
    \end{cases}
\end{eqnarray}
Here $\xi_1$ refers to a radius defined by the condition $\varepsilon_c(\xi_1) = \varepsilon_\mathrm{crit}(\xi_0,\epsilon_\sigma)$. In a sense, it is an equivalent of the so-called marginally bound orbit---a circular orbit of radius
\begin{eqnarray}
     \xi_\mathrm{mb}=2-\epsilon_{\sigma} \alpha + 2 \sqrt{1-\epsilon_{\sigma} \alpha}.
\end{eqnarray}
It has the property that $\varepsilon_c(\xi_\mathrm{mb}) = 1$.

The function $\Lambda_\mathrm{max}(\xi,\xi_0,\varepsilon,\epsilon_\sigma)$ is defined as
\begin{equation}
   \Lambda_\mathrm{max}(\xi,\xi_0,\varepsilon,\epsilon_\sigma) = \mathrm{min}  \left\{ \lambda_\mathrm{max}(\xi,\varepsilon,\epsilon_\sigma), \lambda_\mathrm{max}(\xi_0,\varepsilon,\epsilon_\sigma) \right\},
   \label{Lambdamax}
\end{equation}
ensuring that the trajectories we take into account can reach $\xi = \xi_0$. This is, however, a delicate point of our analysis. A perfectly viable model can also be obtained by setting $\Lambda_\mathrm{max}(\xi,\xi_0,\varepsilon,\epsilon_\sigma) = \lambda_\mathrm{max}(\xi,\varepsilon,\epsilon_\sigma)$. Depending on the energy $\varepsilon$, this choice may correspond to taking into account bounded Keplerian-type orbits with two radial turning points $\xi_1$ and $\xi_2$ satisfying $\xi_1 \le \xi_2 \le \xi_0$. Because our model excludes self-gravity of the gas as well as collisions between individual gas particles, addition of such orbits does not influence the properties of the remaining part of the flow. Neither does it change the accretion rates (see below) or the properties at $\xi = \xi_0$. On the other hand, it can lead to a smooth particle current surface density, as opposed to the choice defined by (\ref{Lambdamax}). In the following, we will show simple examples of the accretion flow obtained by assuming both choices---with and without additional bounded orbits.

\begin{figure}[t]
\subfigure[]{
\label{fig:4a}\includegraphics[width=88mm]{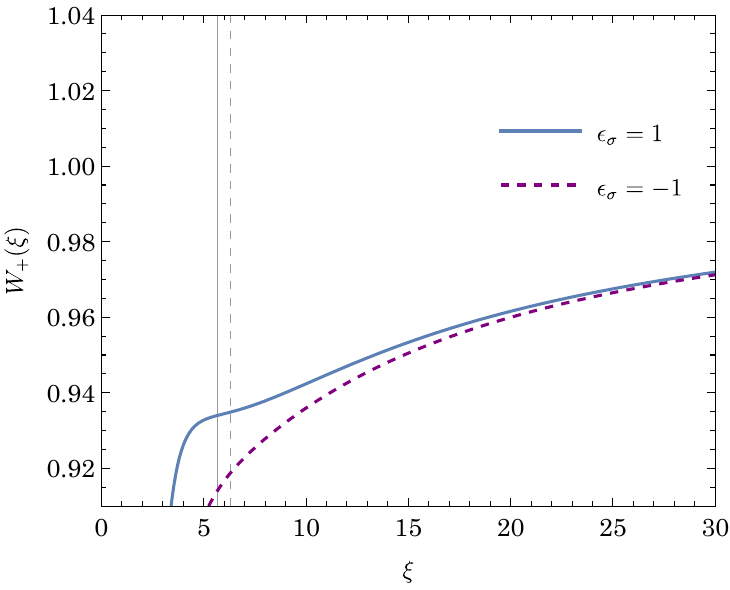}} 
\subfigure[]{
\label{fig:4b}\includegraphics[width=88mm]{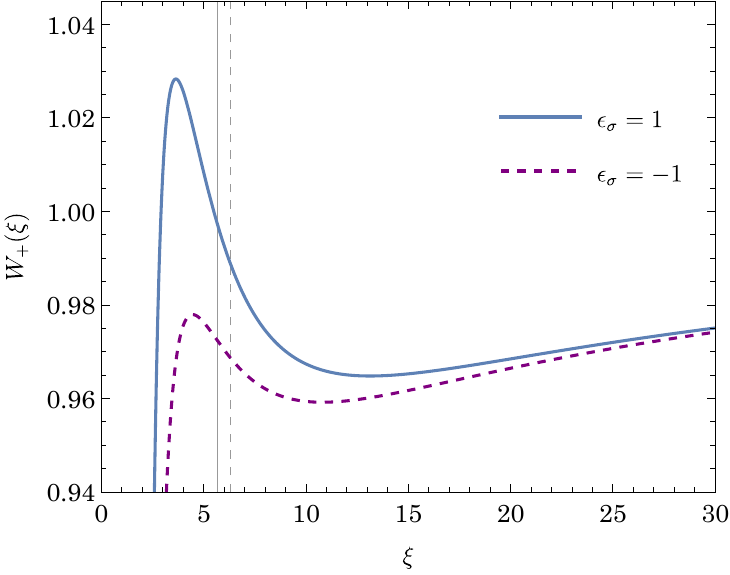}}
\caption{\label{radial potential W_+} Effective potential $W_+(\xi)$. Panel (a): $\lambda = 3.2$. Panel (b): $\lambda = 4$. In both plots $\alpha = 0.12$. Vertical lines mark locations corresponding to marginally stable orbits $\xi_{ms}$.}
\end{figure}

\section{General accretion model}
\label{sec:accretionmodel}

\subsection{Particle current surface density, energy-momentum tensor surface density}

In what follows we consider a gas of the same rest-mass particles characterized by the distribution function depending on $m$, $\varepsilon$, and $\lambda$. We take, in particular
\begin{eqnarray}
    f = A \delta(m - m_0) f_0(\varepsilon,\lambda).
\end{eqnarray}
This leads, for a stationary planar model, to the following expressions for the components of the particle current surface density:
\begin{subequations}
\begin{eqnarray}
\label{Jtgeneral}
    J_t & = & - A m_0^3 \xi \sum_{\epsilon_\sigma = \pm 1} \int \frac{f_0(\varepsilon,\lambda) \varepsilon d \varepsilon d \lambda}{\sqrt{\tilde R}}, \\
    J^r & = & \frac{A m_0^3}{\xi} \sum_{\epsilon_\sigma = \pm 1} \int \epsilon_r f_0(\varepsilon,\lambda) d \varepsilon d \lambda, \\
    J_\varphi & = & A m_0^3 M \xi \sum_{\epsilon_\sigma = \pm 1} \int \frac{f_0(\varepsilon,\lambda) (\epsilon_\sigma \lambda + \alpha \varepsilon) d \varepsilon d \lambda}{\sqrt{\tilde R}}.
\end{eqnarray}
\end{subequations}
In the same fashion, one can obtain the expressions for the components of the energy-momentum surface density. Here we only list expressions for $T\indices{^t_r}$ and $T\indices{^r_\varphi}$, as they will be required to compute the energy and angular momentum accretion rates. They read:
\begin{subequations}
\label{tmunuaccr}
\begin{eqnarray}
T\indices{^r_t} & = & - \frac{A m_0^4}{\xi} \sum_{\epsilon_\sigma = \pm 1} \int \epsilon_r f_0(\varepsilon,\lambda) \varepsilon d \varepsilon d \lambda,\\
T\indices{^r_\varphi} & = & \frac{A M m_0^4}{\xi} \sum_{\epsilon_\sigma = \pm 1} \int \epsilon_r f_0(\varepsilon,\lambda) (\epsilon_\sigma \lambda + \alpha \varepsilon) d \varepsilon d \lambda.
\end{eqnarray}
\end{subequations}

The lack of limits in the integrals in (\ref{Jtgeneral})--(\ref{tmunuaccr}) means that one still has to define the model by specifying the relevant phase-space regions, or in more physical terms, particle orbits which are taken into account. We have already emphasized freedom in this respect, by mentioning the possibility of defining new models by additionally taking into account a collection of Keplerian-type orbits. This also means that in Eqs.\ (\ref{Jtgeneral})--(\ref{tmunuaccr}), as well as in all formulas appearing in the remainder of this paper, we effectively change the convention with respect to the one used in Sec.\ \ref{sec:vlasov} and \ref{sec:vlasovequatorial}, in which the selection of appropriate phase-space ranges was made by a suitable definition of the distribution function.

Specifying the integration limits explicitly, we get for our accretion models
\begin{subequations}
\label{Jmugeneral}
\begin{eqnarray}
        J_t^\mathrm{(abs)} & = & - A m_0^3 \xi \sum_{\epsilon_\sigma = \pm 1} \left[ \int_{W_\mathrm{min}(\xi_0,\epsilon_\sigma)}^{\varepsilon_\mathrm{crit}(\xi_0,\epsilon_\sigma)} d\varepsilon \int_{\Lambda_\mathrm{min}(\xi_0,\varepsilon,\epsilon_\sigma)}^{\lambda_\mathrm{max}(\xi_0,\varepsilon,\epsilon_\sigma)} d \lambda \frac{f_0(\varepsilon,\lambda) \varepsilon}{\sqrt{\tilde R}} +  \int_{\varepsilon_\mathrm{crit}(\xi_0,\epsilon_\sigma)}^\infty d \varepsilon \int_{0}^{\lambda_c(\varepsilon,\epsilon_\sigma)} d \lambda \frac{f_0(\varepsilon,\lambda) \varepsilon}{\sqrt{\tilde R}} \right], \\
    J_t^\mathrm{(scat)} & = & - 2 A m_0^3 \xi \sum_{\epsilon_\sigma = \pm 1} \int_{\varepsilon_\mathrm{min}(\xi,\epsilon_\sigma)}^\infty d \varepsilon \int_{\lambda_c(\varepsilon,\epsilon_\sigma)}^{\Lambda_\mathrm{max}(\xi,\xi_0,\varepsilon,\epsilon_\sigma)} d \lambda \frac{f_0(\varepsilon,\lambda) \varepsilon}{\sqrt{\tilde R}}, \\
    J^r_\mathrm{(abs)} & = & - \frac{A m_0^3}{\xi} \sum_{\epsilon_\sigma = \pm 1}  \left[ \int_{W_\mathrm{min}(\xi_0,\epsilon_\sigma)}^{\varepsilon_\mathrm{crit}(\xi_0,\epsilon_\sigma)} d\varepsilon \int_{\Lambda_\mathrm{min}(\xi_0,\varepsilon,\epsilon_\sigma)}^{\lambda_\mathrm{max}(\xi_0,\varepsilon,\epsilon_\sigma)} d \lambda f_0(\varepsilon,\lambda) +  \int_{\varepsilon_\mathrm{crit}(\xi_0,\epsilon_\sigma)}^\infty d \varepsilon \int_{0}^{\lambda_c(\varepsilon,\epsilon_\sigma)} d \lambda f_0(\varepsilon,\lambda) \right], \label{jrabs} \\
    J^r_\mathrm{(scat)} & = & 0, \\
    J_\varphi^\mathrm{(abs)} & = & A m_0^3 M \xi \sum_{\epsilon_\sigma = \pm 1} \left[ \int_{W_\mathrm{min}(\xi_0,\epsilon_\sigma)}^{\varepsilon_\mathrm{crit}(\xi_0,\epsilon_\sigma)} d \varepsilon \int_{\Lambda_\mathrm{min}(\xi_0,\varepsilon,\epsilon_\sigma)}^{\lambda_\mathrm{max}(\xi_0,\varepsilon,\epsilon_\sigma)} d \lambda \frac{f_0(\varepsilon,\lambda) (\epsilon_\sigma \lambda + \alpha \varepsilon)}{\sqrt{\tilde R}} \right. \nonumber \\
    && \left. + \int_{\varepsilon_\mathrm{crit}(\xi_0,\epsilon_\sigma)}^\infty d \varepsilon \int_{0}^{\lambda_c(\varepsilon,\epsilon_\sigma)} d \lambda \frac{f_0(\varepsilon,\lambda) (\epsilon_\sigma \lambda + \alpha \varepsilon)}{\sqrt{\tilde R}} \right], \\
     J_\varphi^\mathrm{(scat)} & = & 2 A m_0^3 M \xi \sum_{\epsilon_\sigma = \pm 1} \int_{\varepsilon_\mathrm{min}(\xi,\epsilon_\sigma)}^\infty d \varepsilon \int_{\lambda_c(\varepsilon,\epsilon_\sigma)}^{\Lambda_\mathrm{max}(\xi,\xi_0,\varepsilon,\epsilon_\sigma)} d \lambda \frac{f_0(\varepsilon,\lambda) (\epsilon_\sigma \lambda + \alpha \varepsilon)}{\sqrt{\tilde R}}.
\end{eqnarray}
\end{subequations}
The overall factor 2 in formulas for $J_t^\mathrm{(scat)}$ and $J_\varphi^\mathrm{(scat)}$ stems from the fact that both ingoing and outgoing particles contribute to these components.

When comparing models corresponding to different parameters, the normalization constant $A$ may become problematic, as it is difficult to be controled in physical terms. As a remedy, we fix the constant $A$ by specifying the value of the surface mass density $\rho_s$ at the outer boundary of the disk, i.e., at $\xi = \xi_0$. To this end, it is convenient to introduce the following, dimensionless particle current density components:
\begin{equation}
\label{jtilde}
    \tilde J_t = \frac{J_t}{A m_0^3}, \quad \tilde J^r = \frac{J^r}{A m_0^3}, \quad \tilde J_\varphi = \frac{J_\varphi}{A M m_0^3}.
\end{equation}
Equation (\ref{JJ}) now gives
\begin{equation}
\label{rhos}
    \rho_s = A m_0^4 \sqrt{\frac{1}{\xi \tilde \Delta} \left\{ \alpha^2 (2 + \xi) \tilde J_t^2 + \xi^3 \left[ \tilde J_t^2 - \left( \tilde J^r \right)^2 \right] - (\xi - 2) \tilde J_\varphi^2 + 4 \alpha \tilde J_t \tilde J_\varphi \right\}},
\end{equation}
which allows one to express $A$ in terms of $\rho_s$ and the components $\tilde J_t$, $\tilde J^r$, and $\tilde J_\varphi$. As an alternative to $\rho_s$, one could also use $m_0 J^t$. It is easy to see that $J^t$ can be expressed as
\begin{equation}
\label{jt}
    J^t = - \left( 1 + \frac{2}{\xi} + \frac{4}{\tilde \Delta} \right) J_t + \frac{2 \xi}{\tilde \Delta} J^r - \frac{2 \alpha}{M \xi \tilde \Delta} J_\varphi.
\end{equation}
While Eqs.\ (\ref{jtilde}--\ref{jt}) remain general, in practice we will only need to evaluate them at $\xi = \xi_0$.

\subsection{Accretion rates}

Accretion flows are usually characterized by accretion rates of relevant physical quantities. A natural way of defining meaningful accretion rates is related to the existence of conserved vector currents. Our model admits three such quantities. The obvious one is the particle current density $\mathcal J_\mu$. The remaining two are obtained by contracting the energy-momentum tensor with the two Killing vectors $k$ and $\chi$, defined in Eq.\ (\ref{killing}). Specifically, we define $\mathcal J^\mu_{(t)} = \mathcal T\indices{^\mu_\nu}k^\nu$ and $\mathcal J^\mu_{(\varphi)} = \mathcal T\indices{^\mu_\nu}\chi^\nu$. Both currents satisfy $\nabla_\mu \mathcal J^\mu_{(t)} = 0$, $\nabla_\mu \mathcal J^\mu_{(\varphi)} = 0$. The definitions of accretion rates follow from the Stokes theorem, which we write as \cite{Caroll}
\begin{equation}
\int_{\mathcal N} d^4 x \sqrt{|\mathrm{det} \, {}^{(4)}g|} \nabla_\mu \mathcal V^\mu = \int_{\partial \mathcal N} d^{3}y \sqrt{|\mathrm{det} \, {}^{(3)}g|} n_\mu \mathcal V^\mu.
\end{equation}
Here $\mathcal N$ denotes a 4-dimensional region in $\mathcal M$, $\partial N$ is a 3-dimensional boundary of $\mathcal N$, and ${}^{(3)}g$ is the metric induced on $\partial N$. The covariant derivative $\nabla_\mu$ is defined with respect to the metric ${}^{(4)}g$, $\mathcal V^\mu$ is a vector field on $\mathcal M$, and $n^\mu$ is the vector field normal to the boundary $\partial \mathcal N$. For $\nabla_\mu \mathcal V^\mu = 0$, we obtain
\begin{equation}
\label{surfaceterm}
    \int_{\partial \mathcal N} d^{3} y \sqrt{|\mathrm{det} \, {}^{(3)}g|} n_\mu \mathcal V^\mu = 0.
\end{equation}
In the context of our model, we can take $\mathcal N = \{(t,r,\vartheta,\varphi) \colon t_1 \le t \le t_2, r_1 \le r \le r_2, 0 \le \vartheta \le \pi, 0 \le \varphi < 2 \pi \}$, where $t_1$, $t_2$, $r_1$, and $r_2$ are fixed. The metric induced on the surface $r = r_1$ reads
\begin{equation}
    {}^{(3)}g = - dt^2 + (r_1^2 + a^2) \sin^2 \vartheta d \varphi^2 + \rho^2 d \vartheta^2 + \frac{2 M r_1}{\varrho^2} (dt - a \sin^2 \vartheta d \varphi)^2,
\end{equation}
where $\rho^2 = r_1^2 + a^2 \cos^2 \vartheta$. The determinant of this metric is $\mathrm{det} \, {}^{(3)}g = - \Delta \rho^2 \sin^2 \vartheta$, where $\Delta = r_1^2 - 2 M r_1 + a^2$. The unit vector normal to the surface $r = r_1$ has the components $n_\mu = (0,\pm \rho/\sqrt{\Delta},0,0)$ (we assume that $r_1 > r_+$). This gives the term
\begin{equation}
    \int_{r = r_1} d^3 y \sqrt{|\mathrm{det} \, {}^{(3)}g|} n_\mu \mathcal V^\mu = - \int_{r = r_1} dt d\vartheta d \varphi \rho^2 \sin^2 \vartheta \mathcal V^r,
\end{equation}
where we chose the minus sign in the expression for $n_\mu$. For a vector field with values supported on the equatorial plane, it is convenient to define $\mathcal V^r = V^r \delta(\vartheta - \pi/2)/r$. This gives
\begin{equation}
    \int_{r = r_1} d^3 y \sqrt{|\mathrm{det} \, {}^{(3)}g|} n_\mu \mathcal V^\mu = - r_1 \int_{r = r_1, \, \vartheta = \pi/2} dt d \varphi V^r.
\end{equation}
The same result can be also obtained by repeating the above calculation in a setting in which all integration regions are confined to the equatorial plane. This calculation was presented in \cite{CMO2022}.

Defining $J^\mu_{(t)} = T\indices{^\mu_\nu}k^\nu$, $J^\mu_{(\varphi)} = T\indices{^\mu_\nu}\chi^\nu$, so that $\mathcal J^\mu_{(t)}(t,r,\vartheta,\varphi) = J^\mu_{(t)}(t,r,\varphi) \delta(\vartheta - \pi/2)/r$, and $\mathcal J^\mu_{(\varphi)}(t,r,\vartheta,\varphi) = J^\mu_{(\varphi)}(t,r,\varphi) \delta(\vartheta - \pi/2)/r$, we get, setting $V^\mu = m_0 J^\mu, - J^\mu_{(t)}, J^\mu_{(\varphi)}$, three accretion rate measures
\begin{subequations}
\begin{eqnarray}
    \dot M & := &  - r_1 \int_{r = r_1, \, \vartheta = \pi/2} d \varphi \, m_0 J^r = - 2 \pi r_1 m_0 J^r, \\
    \dot {\mathcal E} & := &  + r_1 \int_{r = r_1, \, \vartheta = \pi/2} d \varphi \, J^r_{(t)} = 2 \pi r_1 T\indices{^r_t}, \\
    \dot {\mathcal L} & := &  - r_1 \int_{r = r_1, \, \vartheta = \pi/2} d \varphi \, J^r_{(\varphi)} = - 2 \pi r_1 T\indices{^r_\varphi}.
\end{eqnarray}
\end{subequations}
We will refer to $\dot M$, $\dot {\mathcal E}$, and $\dot {\mathcal L}$, as the mass, the energy, and the angular momentum accretion rates, respectively. The signs in the above definitions were selected in such a way that positive values of $\dot M$, $\dot {\mathcal E}$, and $\dot {\mathcal L}$ would contribute to an increase of the mass and the angular momentum of the black hole.

By construction, for a stationary flow, all accretion rates are independent of the radius $r_1$. Consequently, it is convenient to set $r_1 = r_0$, i.e., compute the accretion rates at the outer boundary of the disk. It is also easy to show that only absorbed trajectories give a net contribution to the accretion rates. Using Eqs.\ (\ref{tmunuaccr}) and (\ref{jrabs}), we get the expressions for the accretion rates in the form
\begin{subequations}
\label{accrrates}
\begin{eqnarray}
\dot M & = & 2 \pi A M m_0^4 \sum_{\epsilon_\sigma = \pm 1}  \left[ \int_{W_\mathrm{min}(\xi_0,\epsilon_\sigma)}^{\varepsilon_\mathrm{crit}(\xi_0,\epsilon_\sigma)} d\varepsilon \int_{\Lambda_\mathrm{min}(\xi_0,\varepsilon,\epsilon_\sigma)}^{\lambda_\mathrm{max}(\xi_0,\varepsilon,\epsilon_\sigma)} d \lambda f_0(\varepsilon,\lambda) +  \int_{\varepsilon_\mathrm{crit}(\xi_0,\epsilon_\sigma)}^\infty d \varepsilon \int_{0}^{\lambda_c(\varepsilon,\epsilon_\sigma)} d \lambda f_0(\varepsilon,\lambda) \right], \\
\dot {\mathcal E} & = & 2 \pi A M m_0^4 \sum_{\epsilon_\sigma = \pm 1}  \left[ \int_{W_\mathrm{min}(\xi_0,\epsilon_\sigma)}^{\varepsilon_\mathrm{crit}(\xi_0,\epsilon_\sigma)} d\varepsilon \varepsilon \int_{\Lambda_\mathrm{min}(\xi_0,\varepsilon,\epsilon_\sigma)}^{\lambda_\mathrm{max}(\xi_0,\varepsilon,\epsilon_\sigma)} d \lambda f_0(\varepsilon,\lambda) +  \int_{\varepsilon_\mathrm{crit}(\xi_0,\epsilon_\sigma)}^\infty d \varepsilon \varepsilon \int_{0}^{\lambda_c(\varepsilon,\epsilon_\sigma)} d \lambda f_0(\varepsilon,\lambda) \right], \\
\dot {\mathcal L} & = & 2 \pi A M^2 m_0^4 \sum_{\epsilon_\sigma = \pm 1}  \left[ \int_{W_\mathrm{min}(\xi_0,\epsilon_\sigma)}^{\varepsilon_\mathrm{crit}(\xi_0,\epsilon_\sigma)} d\varepsilon \int_{\Lambda_\mathrm{min}(\xi_0,\varepsilon,\epsilon_\sigma)}^{\lambda_\mathrm{max}(\xi_0,\varepsilon,\epsilon_\sigma)} d \lambda f_0(\varepsilon,\lambda)(\epsilon_\sigma \lambda + \alpha \varepsilon) \right. \nonumber \\
&& \left. +  \int_{\varepsilon_\mathrm{crit}(\xi_0,\epsilon_\sigma)}^\infty d \varepsilon \int_{0}^{\lambda_c(\varepsilon,\epsilon_\sigma)} d \lambda f_0(\varepsilon,\lambda)(\epsilon_\sigma \lambda + \alpha \varepsilon) \right].
\end{eqnarray}
\end{subequations}

Here, again, it is convenient to express the constant $A$ in terms of the boundary value of the particle surface density $\rho_s$. For $\dot {\mathcal E}$ and $\dot{\mathcal L}$ a more reasonable normalization can be given in terms of the surface energy density $\varepsilon_s = - T\indices{^t_t}$. It can be written in the following form
\begin{eqnarray}
    \varepsilon_s & = & - A  m_0^2 \xi \int f_0(\varepsilon,\lambda) (p^t p_t)_{m = m_0} \frac{d \varepsilon d \lambda}{\sqrt{\tilde R}} \nonumber \\
    & = & A m_0^4 \int f_0 (\varepsilon,\lambda) \varepsilon \left[ \tilde \Delta (\xi + 2) \varepsilon + 2 \left( 2 \xi \varepsilon - \epsilon_\sigma \alpha \lambda - \alpha^2 \varepsilon + \epsilon_r \sqrt{\tilde R} \right) \right] \frac{d\varepsilon d \lambda}{\tilde \Delta \sqrt{\tilde R}}.
\end{eqnarray}
In explicit terms,
\begin{subequations}
\begin{eqnarray}
    \varepsilon_s^\mathrm{(abs)} & = & A m_0^4 \sum_{\epsilon_\sigma = \pm 1} \left\{ \int_{W_\mathrm{min}(\xi_0,\epsilon_\sigma)}^{\varepsilon_\mathrm{crit}(\xi_0,\epsilon_\sigma)} d \varepsilon \int_{\Lambda_\mathrm{min}(\xi_0,\varepsilon,\epsilon_\sigma)}^{\lambda_\mathrm{max}(\xi_0,\varepsilon,\epsilon_\sigma)} d \lambda \frac{f_0 (\varepsilon,\lambda) \varepsilon \left[ \tilde \Delta (\xi + 2) \varepsilon + 2 \left( 2 \xi \varepsilon - \epsilon_\sigma \alpha \lambda - \alpha^2 \varepsilon - \sqrt{\tilde R} \right) \right]}{\tilde \Delta \sqrt{\tilde R}} \right. \nonumber \\
    && \left. + \int_{\varepsilon_\mathrm{crit}(\xi_0,\epsilon_\sigma)}^\infty d \varepsilon \int_{0}^{\lambda_c(\varepsilon,\epsilon_\sigma)} d \lambda \frac{f_0 (\varepsilon,\lambda) \varepsilon \left[ \tilde \Delta (\xi + 2) \varepsilon + 2 \left( 2 \xi \varepsilon - \epsilon_\sigma \alpha \lambda - \alpha^2 \varepsilon - \sqrt{\tilde R} \right) \right]}{\tilde \Delta \sqrt{\tilde R}}  \right\}, \\
    \varepsilon_s^\mathrm{(scat)} & = & 2 A m_0^4 \sum_{\epsilon_\sigma = \pm 1} \int_{\varepsilon_\mathrm{min}(\xi,\epsilon_\sigma)}^\infty d \varepsilon \int_{\lambda_c(\varepsilon,\epsilon_\sigma)}^{\Lambda_\mathrm{max}(\xi,\xi_0,\varepsilon,\epsilon_\sigma)} d \lambda \frac{f_0 (\varepsilon,\lambda) \varepsilon \left[ \tilde \Delta (\xi + 2) \varepsilon + 2 \left( 2 \xi \varepsilon - \epsilon_\sigma \alpha \lambda - \alpha^2 \varepsilon \right) \right]}{\tilde \Delta \sqrt{\tilde R}},
\end{eqnarray}
\end{subequations}
where $\varepsilon_s = \varepsilon_s^\mathrm{(abs)} + \varepsilon_s^\mathrm{(scat)}$.

\section{Accretion of monoenergetic particles}
\label{sec:monoenergetic}

To obtain one of simplest accretion models, one can assume $f_0$ in the form $f_0 = \delta(\varepsilon - \varepsilon_0)$, corresponding to a gas of monoenergetic particles. In this case the planar distribution function reads
\begin{eqnarray}
f = A \delta (m - m_0)\delta (\varepsilon - \varepsilon_0).
\end{eqnarray}
While this choice may seem slightly artificial, it reveals the structure related to the properties of the radial potential $\tilde R$ and the allowed ranges of $\lambda$. For similar reasons, it was also used in \cite{olivierfinite} and in a recent work on Monte Carlo simulations of stationary general-relativistic Vlasov systems \cite{MC1}.

\subsection{Particle current surface density}

\begin{figure}[t]
\subfigure[]{
\includegraphics[width=0.49\textwidth]{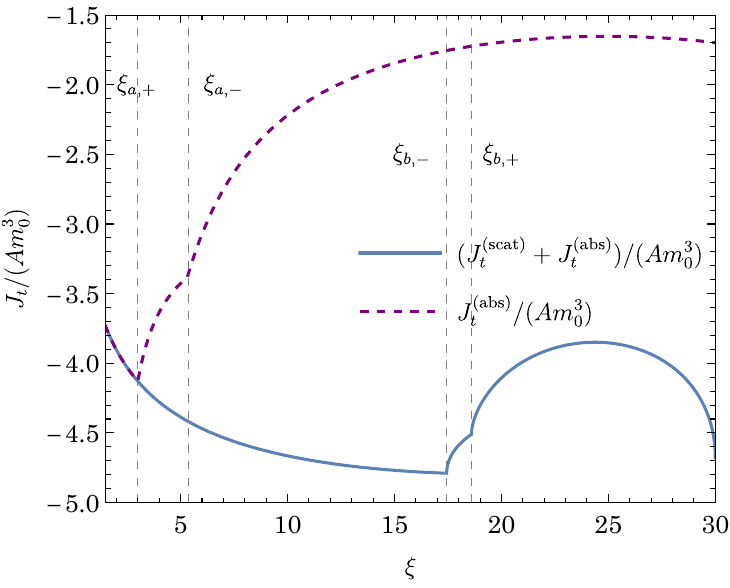}}
\subfigure[]{
\includegraphics[width=0.49\textwidth]{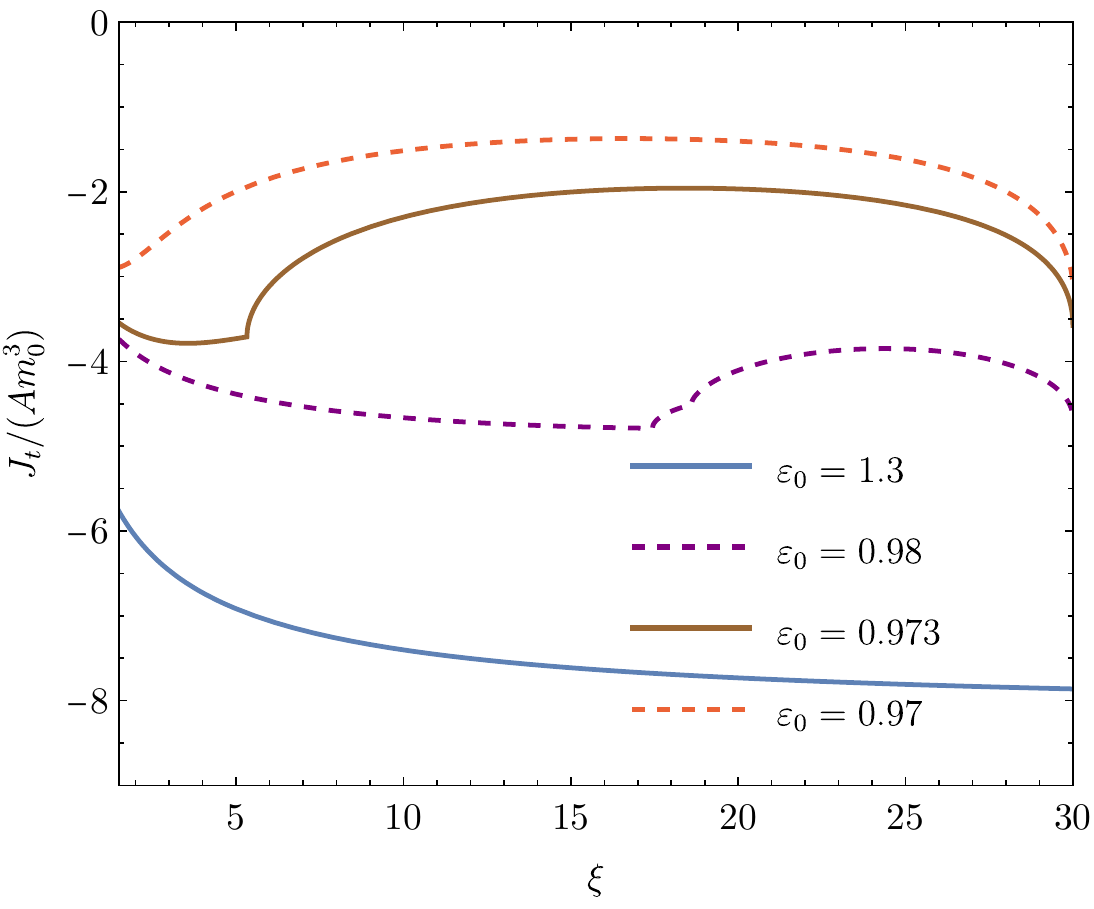}}
\caption{\label{fig:jtmonomin} Time components of the particle current surface density versus radius $\xi$. In all plots $\alpha = 0.5$, $\xi_0 = 30$. Left: An example with $\varepsilon_0 = 0.98$. We plot both $J_t^\mathrm{(abs)}$ and the total component $J_t = J_t^\mathrm{(abs)} + J_t^\mathrm{(scat)}$. Right: Plots of total components $J_t$ for a sample of energies $\varepsilon_0 = 1.3$, 0.98, 0.973, and 0.97.}
\end{figure}

\begin{figure}[t]
\subfigure[]{
\includegraphics[width=0.49\textwidth]{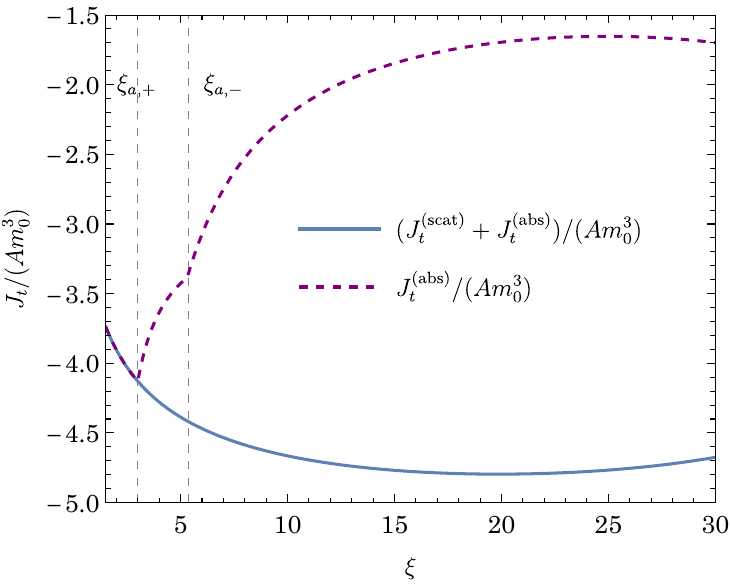}}
\subfigure[]{
\includegraphics[width=0.49\textwidth]{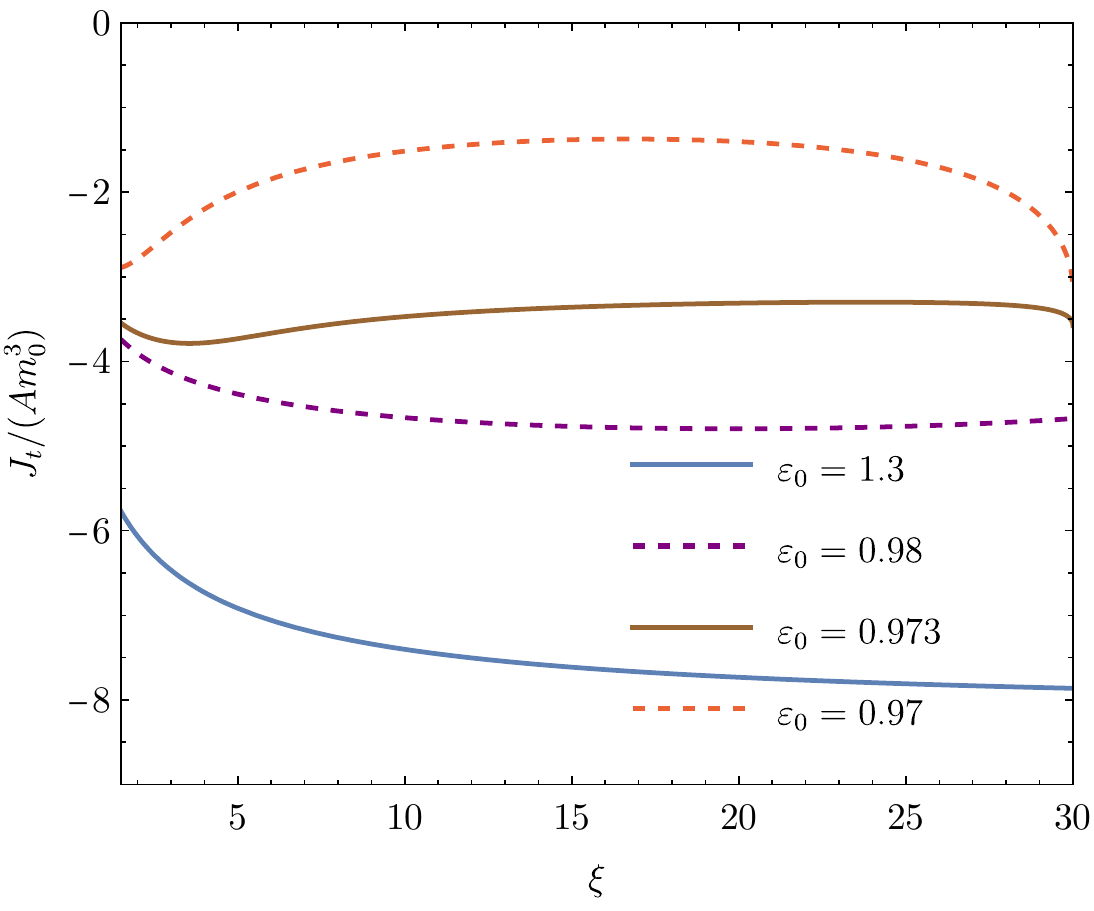}}
\caption{\label{fig:jtmono} Same as in Fig.\ \ref{fig:jtmonomin}, but with $\Lambda_\mathrm{max}(\xi,\xi_0,\varepsilon,\epsilon_\sigma) = \lambda_\mathrm{max}(\xi,\varepsilon,\epsilon_\sigma)$.}
\end{figure}

\begin{figure}[t]
\subfigure[]{
\includegraphics[width=0.49\textwidth]{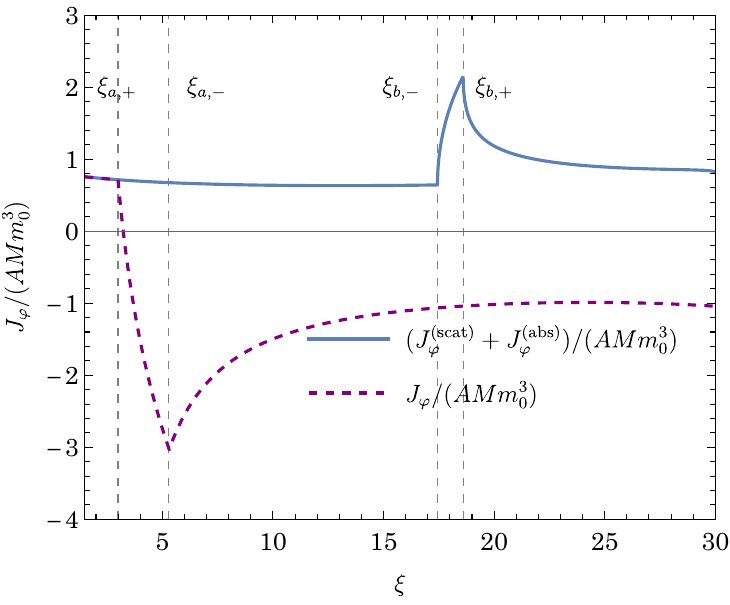}}
\subfigure[]{
\includegraphics[width=0.49\textwidth]{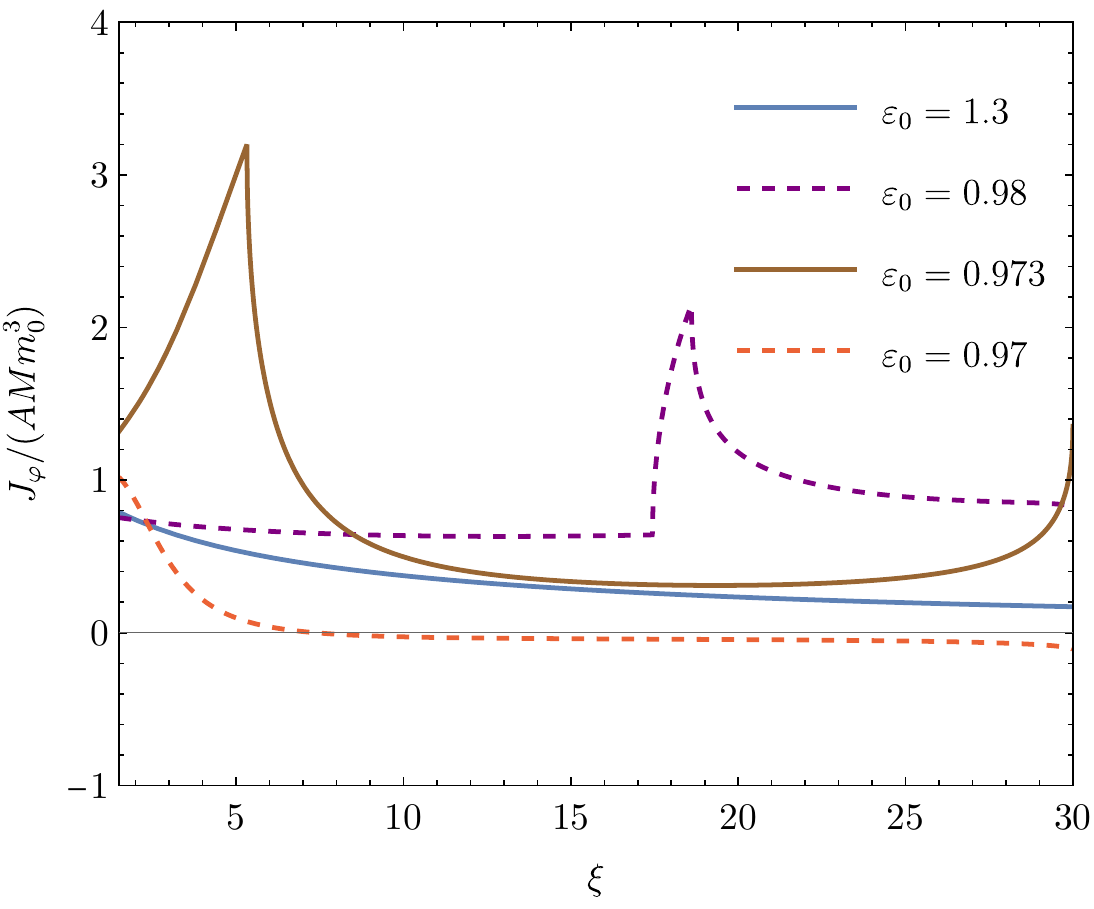}}
\caption{\label{fig:jphimonomin} Angular components of the particle current surface density versus radius $\xi$. In all plots $\alpha = 0.5$, $\xi_0 = 30$. Left: An example with $\varepsilon_0 = 0.98$. We plot both $J_\varphi^\mathrm{(abs)}$ and the total component $J_\varphi = J_\varphi^\mathrm{(abs)} + J_\varphi^\mathrm{(scat)}$. Right: Plots of total components $J_\varphi$ for a sample of energies $\varepsilon_0 = 1.3$, 0.98, 0.973, and 0.97.}
\end{figure}

\begin{figure}[t]
\subfigure[]{
\includegraphics[width=0.49\textwidth]{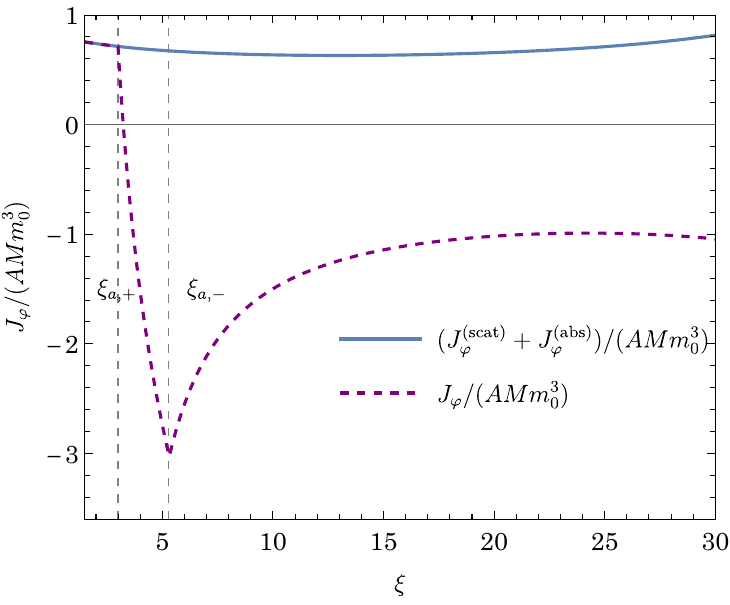}}
\subfigure[]{
\includegraphics[width=0.49\textwidth]{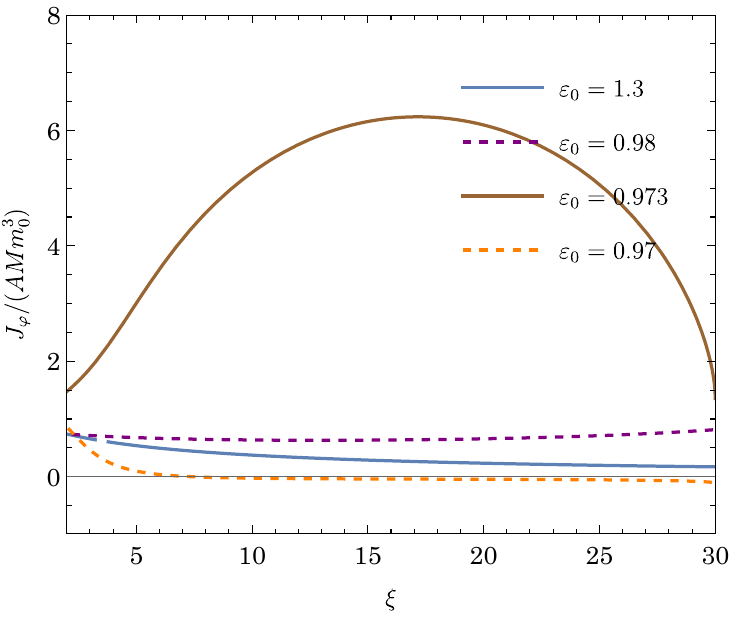}}
\caption{\label{fig:jphimono} Same as in Fig.\ \ref{fig:jphimonomin}, but with $\Lambda_\mathrm{max}(\xi,\xi_0,\varepsilon,\epsilon_\sigma) = \lambda_\mathrm{max}(\xi,\varepsilon,\epsilon_\sigma)$.}
\end{figure}

\begin{figure}[t]
\subfigure[]{
\includegraphics[width=0.49\textwidth]{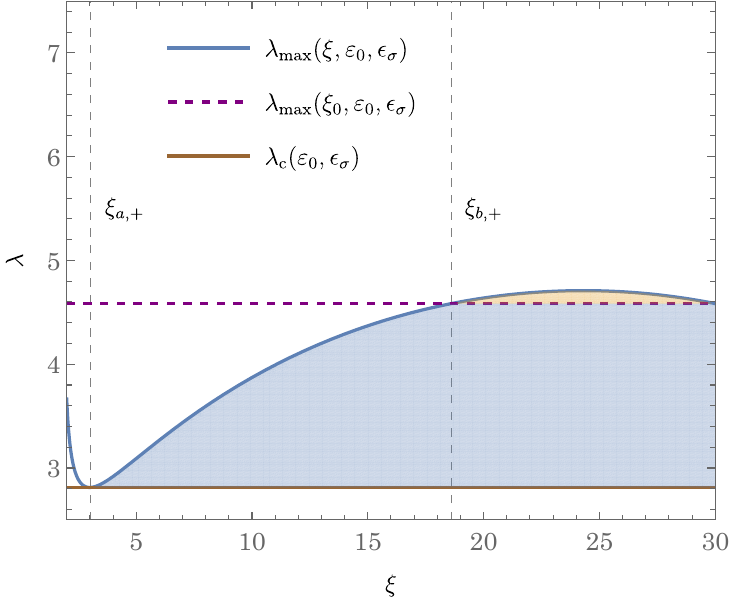}}
\subfigure[]{
\includegraphics[width=0.49\textwidth]{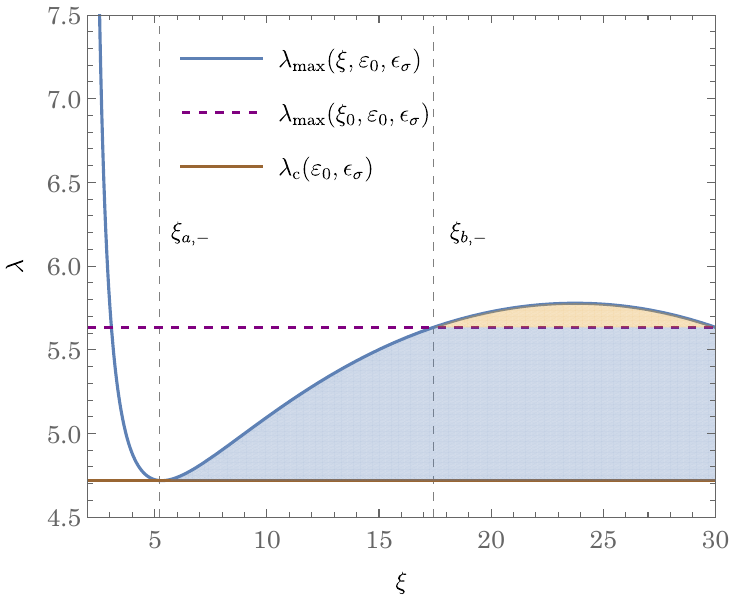}}
\caption{\label{fig:lambdaregions} Ranges of the constant $\lambda$ corresponding to scattered trajectories. Left panel: $\epsilon_\sigma = +1$. Right panel: $\epsilon_\sigma = -1$. The remaining parameters are: $\varepsilon_0 = 0.98$, $\xi_0 = 30$, $\alpha = 1/2$. Blue-shaded regions correspond to $\Lambda_\mathrm{max}(\xi,\xi_0,\varepsilon,\epsilon_\sigma)$ given by Eq.\ (\ref{Lambdamax}). Additional Keplerian-type orbits taken into account by setting instead $\Lambda_\mathrm{max}(\xi,\xi_0,\varepsilon,\epsilon_\sigma) = \lambda_\mathrm{max}(\xi,\varepsilon,\epsilon_\sigma)$ correspond to yellow-shaded regions. None of those orbits reach $\xi = \xi_0$.}
\end{figure}

\begin{figure}[t]
\subfigure[]{
\includegraphics[width=0.49\textwidth]{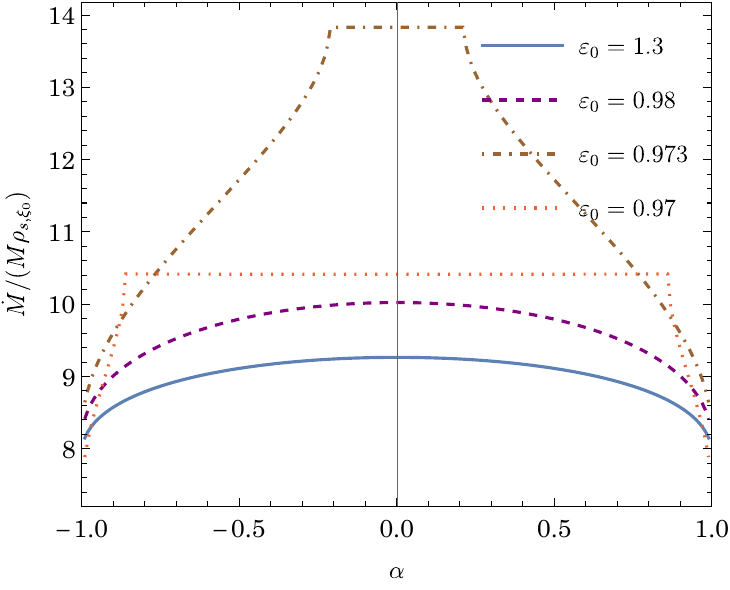}}
\subfigure[]{
\includegraphics[width=0.49\textwidth]{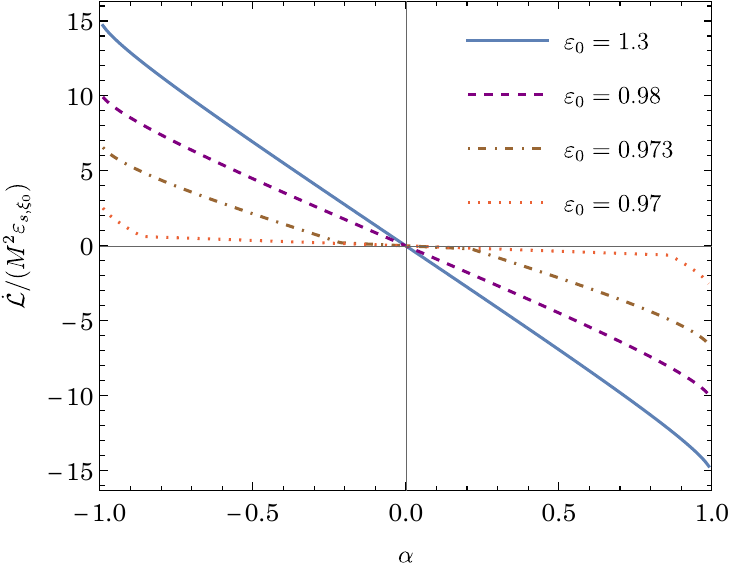}}
\caption{\label{fig:MdotLdotmono} Rest mass and angular momentum accretion rates for monoenergetic models with $\xi_0 = 30$. We plot results for $\varepsilon_0 = 1.3$, 0.98, 0.973, and 0.97.}
\end{figure}

The expressions for the particle current surface density can be now written as follows. For components $J_\mu^\mathrm{(abs)}$, corresponding to absorbed trajectories, we get
\begin{subequations}
\label{e0abs}
\begin{eqnarray}
J^\mathrm{(abs)}_{t}  & = & -A m^3_{0} \xi \varepsilon_{0} \sum_{\epsilon_\sigma = \pm 1} \left [ B(\varepsilon_0,\xi_0,\epsilon_\sigma) \int_{\Lambda_\mathrm{min}(\xi_0,\varepsilon_0,\epsilon_\sigma)}^{\lambda_\mathrm{max}(\xi_0, \varepsilon_0, \epsilon_\sigma)} \frac{d \lambda}{\sqrt{\tilde{R}}} + C(\varepsilon_0,\xi_0,\epsilon_\sigma) \int^{\lambda_{c}(\varepsilon_0,\epsilon_{\sigma})}_{0} \frac{d \lambda}{\sqrt{\tilde{R}}} \right ], \\
J^{r}_\mathrm{(abs)} & = & \frac{-A m^3_{0}}{\xi} \sum_{\epsilon_{\sigma}= \pm 1} \left\{ B(\varepsilon_0,\xi_0,\epsilon_\sigma) \left[ \lambda_\mathrm{max}(\xi_0,\varepsilon_0,\epsilon_\sigma) -\Lambda_\mathrm{min}(\xi_0,\varepsilon_0,\epsilon_\sigma) \right] + C(\varepsilon_0,\xi_0,\epsilon_\sigma) \lambda_c(\varepsilon_0,\epsilon_\sigma) \right\} , \\
   J^\mathrm{(abs)}_\varphi & = & A m^3_{0} M \xi \sum_{\epsilon_{\sigma=\pm 1}} \left[B(\varepsilon_0,\xi_0,\epsilon_\sigma)  \int^{\lambda_\mathrm{max}(\xi_{0},\varepsilon_0,\epsilon_\sigma)}_{\Lambda_\mathrm{min}(\xi_0,\varepsilon_0,\epsilon_\sigma)} d \lambda \frac{\epsilon_\sigma \lambda + \alpha \varepsilon_0}{\sqrt{\tilde{R}}} + C(\varepsilon_0,\xi_0,\epsilon_\sigma) \int^{\lambda_{c}(\varepsilon_0,\epsilon_\sigma)}_{0}  d \lambda \frac{\epsilon_\sigma \lambda + \alpha \varepsilon_0}{\sqrt{\tilde{R}}} \right] \nonumber \\
   & = & A m^3_{0} M \xi \sum_{\epsilon_{\sigma=\pm 1}} \left[ \epsilon_\sigma B(\varepsilon_0,\xi_0,\epsilon_\sigma)  \int^{\lambda_\mathrm{max}(\xi_{0},\varepsilon_0,\epsilon_\sigma)}_{\Lambda_\mathrm{min}(\xi_0,\varepsilon_0,\epsilon_\sigma)}  \frac{\lambda d \lambda}{\sqrt{\tilde{R}}}  + \alpha \varepsilon_0 B(\varepsilon_0,\xi_0,\epsilon_\sigma) \int^{\lambda_\mathrm{max}(\xi_{0},\varepsilon_0,\epsilon_\sigma)}_{\Lambda_\mathrm{min}(\xi_0,\varepsilon_0,\epsilon_\sigma)} \frac{d \lambda}{\sqrt{\tilde{R}}} \right. \nonumber \\
   && \left. +  \epsilon_\sigma C(\varepsilon_0,\xi_0,\epsilon_\sigma) \int^{\lambda_{c}(\varepsilon_0,\epsilon_\sigma)}_{0} \frac{\lambda d\lambda}{\sqrt{\tilde{R}}} + \alpha \varepsilon_0 C(\varepsilon_0,\xi_0,\epsilon_\sigma) \int^{\lambda_{c}(\varepsilon_0,\epsilon_\sigma)}_{0} \frac{d \lambda}{\sqrt{\tilde{R}}}\right],
\end{eqnarray}
\end{subequations}
where we have defined
\begin{subequations}
\label{BandC}
\begin{eqnarray}
B(\varepsilon_0,\xi_0,\epsilon_\sigma) & := & \chi ( W_\mathrm{min}(\xi_0,\epsilon_\sigma) \le \varepsilon_0 < \varepsilon_\mathrm{crit}(\xi_0,\epsilon_\sigma) ), \\
C(\varepsilon_0,\xi_0,\epsilon_\sigma) & := & \chi(\varepsilon_\mathrm{crit}(\xi_0,\epsilon_\sigma) \le \varepsilon_0),
\end{eqnarray}
\end{subequations}
and $\chi$ denotes the characteristic function ($\chi(\mathcal{C})$ equals $1$ if condition $\mathcal{C}$ is satisfied and $0$ otherwise).
The expressions for $J_\mu^\mathrm{(scat)}$ can be written as
\begin{subequations}
\label{e0scat}
\begin{eqnarray}
J^\mathrm{(scat)}_{t} & = & -2 A m^3_{0}  \xi \varepsilon_{0} \sum_{\epsilon_{\sigma=\pm 1}}  D(\varepsilon_0,\xi,\epsilon_\sigma) \int^{\Lambda_\mathrm{max}(\xi,\xi_0,\varepsilon_0,\epsilon_{\sigma})}_{\lambda_{c}(\varepsilon_0,\epsilon_{\sigma})} \frac{d \lambda}{\sqrt{\tilde{R}}},\\
J^\mathrm{(scat)}_{r} & = & 0, \\
J^\mathrm{(scat)}_{\varphi} & = & 2 A m^3_{0} M \xi \sum_{\epsilon_{\sigma=\pm 1}} D(\varepsilon_0,\xi,\epsilon_\sigma)  \int^{\Lambda_\mathrm{max}(\xi,\xi_0,\varepsilon_0,\epsilon_{\sigma})}_{\lambda_{c}(\varepsilon_0,\epsilon_{\sigma})} d \lambda \frac{\epsilon_\sigma \lambda + \alpha \varepsilon_{0}}{\sqrt{\tilde{R}}}  \nonumber \\
& = & 2 A m^3_{0} M \xi \sum_{\epsilon_{\sigma=\pm 1}} \left[ \epsilon_\sigma D(\varepsilon_0,\xi,\epsilon_\sigma)  \int^{\Lambda_\mathrm{max}(\xi,\xi_0,\varepsilon_0,\epsilon_{\sigma})}_{\lambda_{c}(\varepsilon_0,\epsilon_{\sigma})} \frac{\lambda d \lambda}{\sqrt{\tilde{R}}} + \alpha \varepsilon_0 D(\varepsilon_0,\xi,\epsilon_\sigma) \int^{\Lambda_\mathrm{max}(\xi,\xi_0,\varepsilon_0,\epsilon_{\sigma})}_{\lambda_{c}(\varepsilon_0,\epsilon_{\sigma})} \frac{d \lambda}{\sqrt{\tilde{R}}} \right],
\end{eqnarray}
\end{subequations}
where
\begin{equation}
    D(\varepsilon_0,\xi,\epsilon_\sigma) = \chi (\varepsilon_\mathrm{min}(\xi,\epsilon_\sigma) \le \varepsilon_0).
\end{equation}
The additional factor of 2 appearing in the expressions for $J_{t}^\mathrm{(scat)}$ and $J_{\varphi}^\mathrm{(scat)}$ arises from the two possible radial directions of motion along a scattered trajectory, denoted as $\epsilon_r = \pm 1$.

Similarly, the expressions for the energy surface density can be written as
\begin{subequations}
\label{monoenergetic_epsilons}
\begin{eqnarray}
    \varepsilon_s^\mathrm{(abs)} & = & A m_0^4 \varepsilon_{0} \sum_{\epsilon_\sigma = \pm 1} \left\{ B(\varepsilon_0,\xi_0,\epsilon_\sigma) \int_{\Lambda_\mathrm{min}(\xi_0,\varepsilon_{0},\epsilon_\sigma)}^{\lambda_\mathrm{max}(\xi_0,\varepsilon_{0},\epsilon_\sigma)} d \lambda \frac{ \tilde \Delta (\xi + 2) \varepsilon_{0} + 2 \left( 2 \xi \varepsilon_{0} - \epsilon_\sigma \alpha \lambda - \alpha^2 \varepsilon_{0} - \sqrt{\tilde R} \right)}{\tilde \Delta \sqrt{\tilde R}} \right. \nonumber \\
    && \left. + C(\varepsilon_0,\xi_0,\epsilon_\sigma) \int_{0}^{\lambda_c(\varepsilon_{0},\epsilon_\sigma)} d \lambda \frac{\tilde \Delta (\xi + 2) \varepsilon_{0} + 2 \left( 2 \xi \varepsilon_{0} - \epsilon_\sigma \alpha \lambda - \alpha^2 \varepsilon_{0} - \sqrt{\tilde R} \right)}{\tilde \Delta \sqrt{\tilde R}} \right\}, \\
    \varepsilon_s^\mathrm{(scat)} & = & 2 A m_0^4  \varepsilon_{0} \sum_{\epsilon_\sigma = \pm 1} D(\varepsilon_0,\xi,\epsilon_\sigma) \int_{\lambda_c(\varepsilon_{0},\epsilon_\sigma)}^{\Lambda_\mathrm{max}(\xi,\xi_0,\varepsilon_{0},\epsilon_\sigma)} d \lambda \frac{\tilde \Delta (\xi + 2) \varepsilon_{0} + 2 \left( 2 \xi \varepsilon_{0} - \epsilon_\sigma \alpha \lambda - \alpha^2 \varepsilon_{0} \right)}{\tilde \Delta \sqrt{\tilde R}}. 
\end{eqnarray}
\end{subequations}
In this work, Eqs.\ (\ref{monoenergetic_epsilons}) are only evaluated at $\xi = \xi_0$ and used to provide a suitable normalization, i.e., to remove the constant $A$, where it is needed.

When evaluating the components of $J_\mu$ at the outer edge of the disk $\xi = \xi_0$, it is convenient to note the following simplifications. For $\xi > 2$, $\epsilon_\sigma = -1$, and $W_\mathrm{min}(\xi_0,\epsilon_\sigma)< \varepsilon < \sqrt{\tilde \Delta(\xi_0)}/\xi_0$, we have
\begin{equation}
    \int_{\lambda_\mathrm{min}(\xi_0,\varepsilon,\epsilon_\sigma)}^{\lambda_\mathrm{max}(\xi_0,\varepsilon,\epsilon_\sigma)} \frac{d \lambda}{\sqrt{\tilde R(\xi_0)}} = \frac{\pi}{\sqrt{\xi_0 (\xi_0 - 2)}}.
\end{equation}
This result does not depend on the value of the energy $\varepsilon$; it stems directly from the factorization
\begin{equation}
    \tilde R(\xi_0) = - \xi_0 (\xi_0 - 2)\left[ \lambda -  \lambda_\mathrm{max}(\xi_0,\varepsilon,\epsilon_\sigma) \right] \left[\lambda- \lambda_\mathrm{min}(\xi_0,\varepsilon,\epsilon_\sigma) \right].
\end{equation}
By an analogous reasoning, we have
\begin{equation}
    \sum_{\epsilon_\sigma = \pm 1} \int_0^{\lambda_\mathrm{max}(\xi_0,\varepsilon,\epsilon_\sigma)} \frac{d \lambda}{\sqrt{\tilde R(\xi_0)}} = \frac{\pi}{\sqrt{\xi_0 (\xi_0 - 2)}},
\end{equation}
for $\varepsilon > \sqrt{\tilde \Delta(\xi_0)}/\xi_0$.

All numerical calculationss presented in this article were performed using \textit{Wolfram Mathematica} \cite{wolfram}. Sample plots of $J_t$ and $J_\varphi$ for monoenergetic models are shown in Figs.\ \ref{fig:jtmonomin}--\ref{fig:jphimono}. In left panels dashed lines are used to plot the components associated with absorbed trajectories. Total components---corresponding to scattered and absorbed trajectories---are plotted with solid lines. In the right panels, we show the dependence of $J_t$ and $J_\varphi$ on the energy $\varepsilon_0$. In all cases, the components of the particle current surface density are normalized by a factor proportional to $A$. The components plotted in Figs.\ \ref{fig:jtmonomin} and \ref{fig:jphimonomin} are computed assuming $\Lambda_\mathrm{max}(\xi,\xi_0,\varepsilon,\epsilon_\sigma)$ defined by Eq.\ (\ref{Lambdamax}). That means that all scattered trajectories reach $\xi = \xi_0$. In contrast to this, Figs.\ \ref{fig:jtmono} and \ref{fig:jphimono} are prepared assuming $\Lambda_\mathrm{max}(\xi,\xi_0,\varepsilon,\epsilon_\sigma) = \lambda_\mathrm{max}(\xi,\varepsilon,\epsilon_\sigma)$, which amounts to adding a suitable set of bounded trajectories to the system.

In all plots, we assume the same values $\alpha = 1/2$ and $\xi_0 = 30$. Different energies $\varepsilon_0 = 0.97$, 0.973, 0.98, and 1.3, are selected to illustrate various possible behaviors of scattered trajectories. For $\alpha = 1/2$ and $\xi_0 = 30$, we have $\varepsilon_\mathrm{crit} (\xi_0,\epsilon_\sigma) = 0.971885$ for $\epsilon_\sigma = +1$ and $\varepsilon_\mathrm{crit} (\xi_0,\epsilon_\sigma) = 0.975229$ for $\epsilon_\sigma = -1$. Thus, $0.97 < \varepsilon_\mathrm{crit} (\xi_0,\epsilon_\sigma=+1) < 0.973 < \varepsilon_\mathrm{crit} (\xi_0,\epsilon_\sigma=-1) < 0.98 < 1.3$. Consequently, the solution for $\varepsilon_0 = 0.97$ involves only absorbed trajectories. For $\varepsilon_0 = 0.973$, there is a component of absorbed trajectories with $\epsilon_\sigma = +1$. For $\varepsilon_0 = 0.98$ scattered trajectories are allowed for both $\epsilon_\sigma = +1$ and $\epsilon_\sigma = -1$. The value $\varepsilon_0 = 1.3$ was chosen to correspond to unbound orbits.

The difference between the choices of $\Lambda_\mathrm{max}(\xi,\xi_0,\varepsilon,\epsilon_\sigma)$ defined by Eq.\ (\ref{Lambdamax}) and $\Lambda_\mathrm{max}(\xi,\xi_0,\varepsilon,\epsilon_\sigma) = \lambda_\mathrm{max}(\xi,\varepsilon,\epsilon_\sigma)$ is illustrated in Fig.\ \ref{fig:lambdaregions}, in which we plot the ranges of $\lambda$ corresponding to scattered trjectories in a sample model with $\alpha = 1/2$, $\xi_0 = 30$, and $\varepsilon_0 = 0.98$. The ranges corresponding to $\Lambda_\mathrm{max}(\xi,\xi_0,\varepsilon,\epsilon_\sigma)$ defined by Eq.\ (\ref{Lambdamax}) are shown in blue shade. Additional Keplerian-type orbits correspond to yellow-shaded regions. Figure \ref{fig:lambdaregions} also allows us to explain the lack of smoothness observed in the plots of $J_t$ or $J_\varphi$. Vertical lines in Fig.\ \ref{fig:lambdaregions} mark the locations characterized by $\lambda_\mathrm{max}(\xi,\varepsilon_0,\epsilon_\sigma) = \lambda_\mathrm{max}(\xi_0,\varepsilon_0,\epsilon_\sigma)$ (they are denoted as $\xi_{b,\pm}$, where the sign $\pm$ corresponds to $\epsilon_\sigma = \pm 1$) and $\lambda_\mathrm{max}(\xi,\varepsilon_0,\epsilon_\sigma) = \lambda_c(\varepsilon_0,\epsilon_\sigma)$ (denoted as $\xi_{a,\pm}$). According to Eq.\ (\ref{scattcondition}), the latter location introduces a cutoff-type behavior: no scattered trajectories (with fixed $\varepsilon_0$ and $\epsilon_\sigma$) reach below that distance. The same four lines (at $\xi_{a,\pm}$ and $\xi_{b,\pm}$) are also shown in the left panels of Figs.\ \ref{fig:jtmonomin}, \ref{fig:jphimonomin}. In addition, locations of $\xi_{a,\pm}$ are plotted in Figs.\ \ref{fig:jtmono} and \ref{fig:jphimono}.

\subsection{Accretions Rates}

For the monoenergetic distribution, accretion rates $\dot M$, $\dot{\mathcal E}$, and $\dot{\mathcal L}$ can be computed exactly (save for the functions $\varepsilon_\mathrm{crit}(\xi_0,\epsilon_\sigma)$ and $\lambda_c(\varepsilon_0,\epsilon_\sigma)$, which still need to be evaluated numerically).
Assuming $f_0 = \delta(\varepsilon - \varepsilon_0)$ in Eqs.\ (\ref{accrrates}), we get
\begin{subequations}
\begin{eqnarray}
    \dot M & = & 2 \pi A M m_0^4 \sum_{\epsilon_\sigma = \pm 1} \left\{ B(\varepsilon_0,\xi_0,\epsilon_\sigma) [\lambda_\mathrm{max}(\xi_0,\varepsilon_0,\epsilon_\sigma)-\Lambda_\mathrm{min}(\xi_0,\varepsilon_0,\epsilon_\sigma)] + C(\varepsilon_0,\xi_0,\epsilon_\sigma) \lambda_c(\varepsilon_0,\epsilon_\sigma) \right\}, \\
    \dot {\mathcal E} & = & \varepsilon_0 \dot M, \\
   \dot{\mathcal{L}} &=&  2 \pi A M^2 m_0^4 \sum_{\epsilon_\sigma = \pm 1}  \Bigg \{ B(\varepsilon_0,\xi_0,\epsilon_\sigma) \left[ \frac{\epsilon_{\sigma}}{2} \left[\lambda^{2}_\mathrm{max}(\xi_0,\varepsilon_0,\epsilon_\sigma) -\Lambda^{2}_\mathrm{min}(\xi_0,\varepsilon_0,\epsilon_\sigma)\right] \right. \nonumber \\
   && \left. \left. +  \alpha \varepsilon_0 \left[\lambda_\mathrm{max}(\xi_0,\varepsilon_0,\epsilon_\sigma) -\Lambda_\mathrm{min}(\xi_0,\varepsilon_0,\epsilon_\sigma)\right] \right] + C(\varepsilon_0,\xi_0,\epsilon_\sigma) \left[ \frac{\epsilon_\sigma}{2} \lambda_c(\varepsilon_0,\epsilon_\sigma) + \alpha \varepsilon_0 \right] \lambda_c(\varepsilon_0,\epsilon_\sigma) \right. \Bigg \}, \label{Ldotmono}
\end{eqnarray}
\end{subequations}
where $B(\varepsilon_0,\xi_0,\epsilon_\sigma)$ and $C(\varepsilon_0,\xi_0,\epsilon_\sigma)$ are defined in Eqs.\ (\ref{BandC}).
Note that $\dot{ \mathcal L} = 0$ for $\alpha = 0$ (a non-rotating black hole), as expected.

Sample plots of $\dot M$ and $\dot{\mathcal L}$ versus $\alpha$ are shown in Fig.\ \ref{fig:MdotLdotmono}. For a fair comparison, we normalize $\dot M$ and $\dot{\mathcal L}$ by the boundary values of the mass and energy surface densities, respectively. They are calculated directly from Eqs.\ (\ref{rhos}) and (\ref{monoenergetic_epsilons}) evaluated at $\xi = \xi_0$. This normalization allows one to remove the proportionality constant $A$, which is otherwise hard to compare between different models. The dependence of $\dot M$ and $\dot{\mathcal L}$ can be quite intricate for energies $\varepsilon_0$ within the range $[W_\mathrm{min}(\xi_0,\epsilon_\sigma),\varepsilon_\mathrm{crit}(\xi_0,\epsilon_\sigma)]$. For larger values of $\varepsilon_0$, one recovers the behavior known from the Maxwell-J\"{u}ttner model analyzed in \cite{CMO2022}. The mass accretion rate $\dot M$ decreases with $|\alpha|$, while the angular momentum accretion rate decreases with $\alpha$. The latter means, in particular, that the black hole spin decreases in a quasi-stationary scenario characterized by the Eq.\ (\ref{Ldotmono}).

\section{Maxwell-J\"{u}ttner type distribution}
\label{sec:maxwelljuttner}

Another choice for $f_0(\varepsilon,\lambda)$, physically more sound than the monoenergetic distribution, is to set
\begin{equation}
\label{MJdist}
    f_0(\varepsilon,\lambda) = \exp \left( -\beta \varepsilon \right),
\end{equation}
where $\beta$ is a constant. This choice corresponds to the so-called Maxwell-J\"{u}ttner distribution \cite{juttner1,juttner2}. The Maxwell-J\"{u}ttner distribution characterizes a gas in thermal equilibrium in the flat Minkowski spacetime. In this case $\beta = m_0/(k_\mathrm{B}T)$, where $T$ denotes the temperature, and $k_\mathrm{B}$ is the Boltzmann constant. In our model, the spacetime at $\xi = \xi_0$ is not flat, and consequently one cannot speak about thermal equilibrium in a strict sense. In addition, we consider the gas to be confined to the equatorial plane. While a two-dimensional Maxwell-J\"{u}ttner distribution still makes sense, its physical properties differ in some aspects from the standard three-dimensional version (see, e.g.\ \cite{intro}).

\subsection{Particle current surface density}

The expressions for the particle current surface density follow directly from Eqs.\ (\ref{Jmugeneral}). Assuming Eq.\ (\ref{MJdist}), we get
\begin{subequations}
\begin{eqnarray}
    J_t^\mathrm{(abs)} & = & - A m_0^3 \xi \sum_{\epsilon_\sigma = \pm 1} \left[ \int_{W_\mathrm{min}(\xi_0,\epsilon_\sigma)}^{\varepsilon_\mathrm{crit}(\xi_0,\epsilon_\sigma)} d\varepsilon \exp(-\beta \varepsilon) \varepsilon \int_{\Lambda_\mathrm{min}(\xi_0,\varepsilon,\epsilon_\sigma)}^{\lambda_\mathrm{max}(\xi_0,\varepsilon,\epsilon_\sigma)}  \frac{d \lambda}{\sqrt{\tilde R}} +  \int_{\varepsilon_\mathrm{crit}(\xi_0,\epsilon_\sigma)}^\infty d \varepsilon \exp(-\beta \varepsilon) \varepsilon \int_{0}^{\lambda_c(\varepsilon,\epsilon_\sigma)} \frac{d \lambda}{\sqrt{\tilde R}} \right], \\
    J_t^\mathrm{(scat)} & = & - 2 A m_0^3 \xi \sum_{\epsilon_\sigma = \pm 1} \int_{\varepsilon_\mathrm{min}(\xi,\epsilon_\sigma)}^\infty d \varepsilon \exp(-\beta \varepsilon) \varepsilon \int_{\lambda_c(\varepsilon,\epsilon_\sigma)}^{\Lambda_\mathrm{max}(\xi,\xi_0,\varepsilon,\epsilon_\sigma)} \frac{d \lambda}{\sqrt{\tilde R}}, \\
    J^r_\mathrm{(abs)} & = & - \frac{A m_0^3}{\xi} \sum_{\epsilon_\sigma = \pm 1}  \left\{ \int_{W_\mathrm{min}(\xi_0,\epsilon_\sigma)}^{\varepsilon_\mathrm{crit}(\xi_0,\epsilon_\sigma)} d\varepsilon \exp(- \beta \varepsilon) \left[ \lambda_\mathrm{max}(\xi_0,\varepsilon,\epsilon_\sigma) - \Lambda_\mathrm{min}(\xi_0,\varepsilon,\epsilon_\sigma) \right] \right. \nonumber \\
    && \left. +  \int_{\varepsilon_\mathrm{crit}(\xi_0,\epsilon_\sigma)}^\infty d \varepsilon \exp(- \beta \varepsilon) \lambda_c(\varepsilon,\epsilon_\sigma)  \right\}, \\
    J^r_\mathrm{(scat)} & = & 0, \\
    J_\varphi^\mathrm{(abs)} & = & A m_0^3 M \xi \sum_{\epsilon_\sigma = \pm 1} \left[ \int_{W_\mathrm{min}(\xi_0,\epsilon_\sigma)}^{\varepsilon_\mathrm{crit}(\xi_0,\epsilon_\sigma)} d \varepsilon \int_{\Lambda_\mathrm{min}(\xi_0,\varepsilon,\epsilon_\sigma)}^{\lambda_\mathrm{max}(\xi_0,\varepsilon,\epsilon_\sigma)} d \lambda \frac{\exp(-\beta \varepsilon)(\epsilon_\sigma \lambda + \alpha \varepsilon)}{\sqrt{\tilde R}} \right. \nonumber \\
    && \left. + \int_{\varepsilon_\mathrm{crit}(\xi_0,\epsilon_\sigma)}^\infty d \varepsilon \int_{0}^{\lambda_c(\varepsilon,\epsilon_\sigma)} d \lambda \frac{\exp(-\beta \varepsilon) (\epsilon_\sigma \lambda + \alpha \varepsilon)}{\sqrt{\tilde R}} \right], \\
     J_\varphi^\mathrm{(scat)} & = & 2 A m_0^3 M \xi \sum_{\epsilon_\sigma = \pm 1} \int_{\varepsilon_\mathrm{min}(\xi,\epsilon_\sigma)}^\infty d \varepsilon \exp(-\beta \varepsilon) \int_{\lambda_c(\varepsilon,\epsilon_\sigma)}^{\Lambda_\mathrm{max}(\xi,\xi_0,\varepsilon,\epsilon_\sigma)} d \lambda \frac{\epsilon_\sigma \lambda + \alpha \varepsilon}{\sqrt{\tilde R}}.
\end{eqnarray}
\end{subequations}
The expression for the energy surface density can be written in the following form:
\begin{subequations}
\begin{eqnarray}
    \varepsilon_s^\mathrm{(abs)} & = & A m_0^4 \sum_{\epsilon_\sigma = \pm 1} \left\{ \int_{W_\mathrm{min}(\xi_0,\epsilon_\sigma)}^{\varepsilon_\mathrm{crit}(\xi_0,\epsilon_\sigma)} d \varepsilon \int_{\Lambda_\mathrm{min}(\xi_0,\varepsilon,\epsilon_\sigma)}^{\lambda_\mathrm{max}(\xi_0,\varepsilon,\epsilon_\sigma)} d \lambda \frac{\exp (-\beta \varepsilon) \varepsilon \left[ \tilde \Delta (\xi + 2) \varepsilon + 2 \left( 2 \xi \varepsilon - \epsilon_\sigma \alpha \lambda - \alpha^2 \varepsilon - \sqrt{\tilde R} \right) \right]}{\tilde \Delta \sqrt{\tilde R}} \right. \nonumber \\
    && \left. + \int_{\varepsilon_\mathrm{crit}(\xi_0,\epsilon_\sigma)}^\infty d \varepsilon \int_{0}^{\lambda_c(\varepsilon,\epsilon_\sigma)} d \lambda \frac{\exp (-\beta \varepsilon) \varepsilon \left[ \tilde \Delta (\xi + 2) \varepsilon + 2 \left( 2 \xi \varepsilon - \epsilon_\sigma \alpha \lambda - \alpha^2 \varepsilon - \sqrt{\tilde R} \right) \right]}{\tilde \Delta \sqrt{\tilde R}} \right\}, \\
    \varepsilon_s^\mathrm{(scat)} & = & 2 A m_0^4 \sum_{\epsilon_\sigma = \pm 1} \int_{\varepsilon_\mathrm{min}(\xi,\epsilon_\sigma)}^\infty d \varepsilon \int_{\lambda_c(\varepsilon,\epsilon_\sigma)}^{\Lambda_\mathrm{max}(\xi,\xi_0,\varepsilon,\epsilon_\sigma)} d \lambda \frac{\exp (-\beta \varepsilon) \varepsilon \left[ \tilde \Delta (\xi + 2) \varepsilon + 2 \left( 2 \xi \varepsilon - \epsilon_\sigma \alpha \lambda - \alpha^2 \varepsilon \right) \right]}{\tilde \Delta \sqrt{\tilde R}}.
    \label{juttner _epsilons}
\end{eqnarray}
\end{subequations}

\begin{figure}[t]
\subfigure[]{
\includegraphics[width=0.49\textwidth]{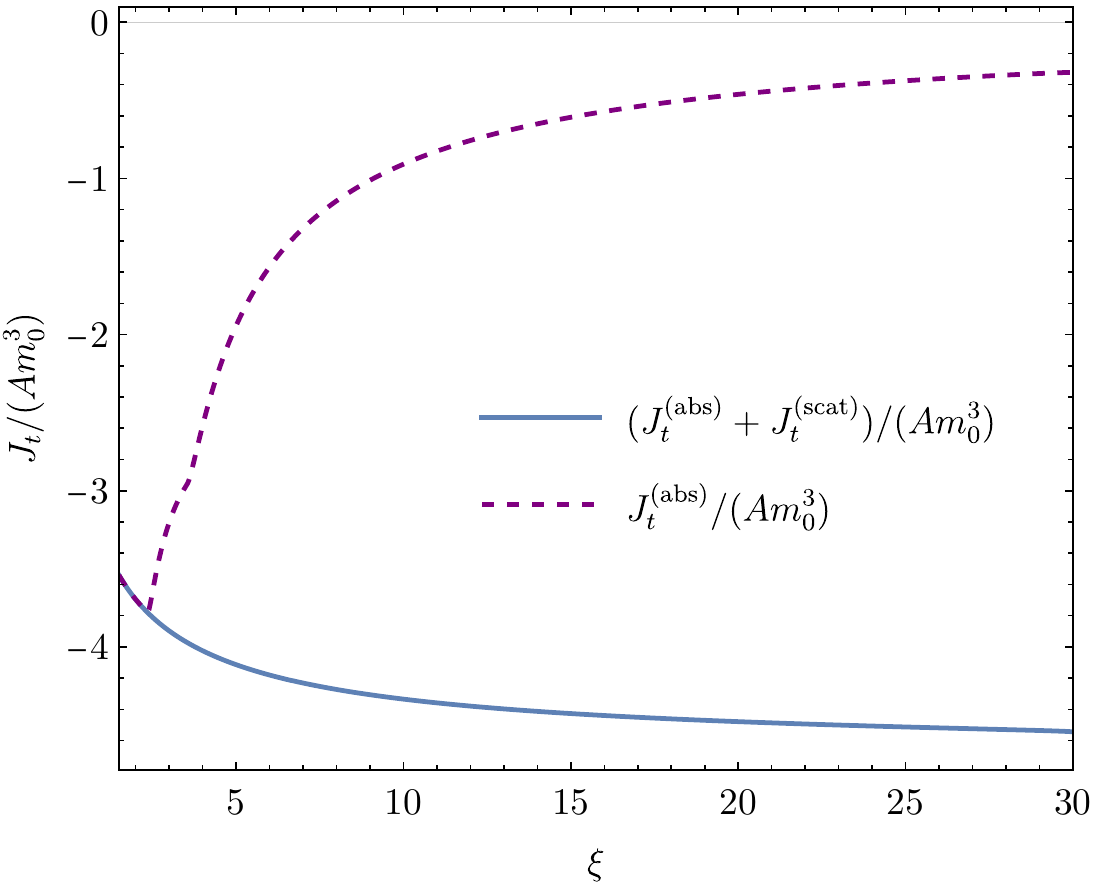}}
\subfigure[]{
\includegraphics[width=0.49\textwidth]{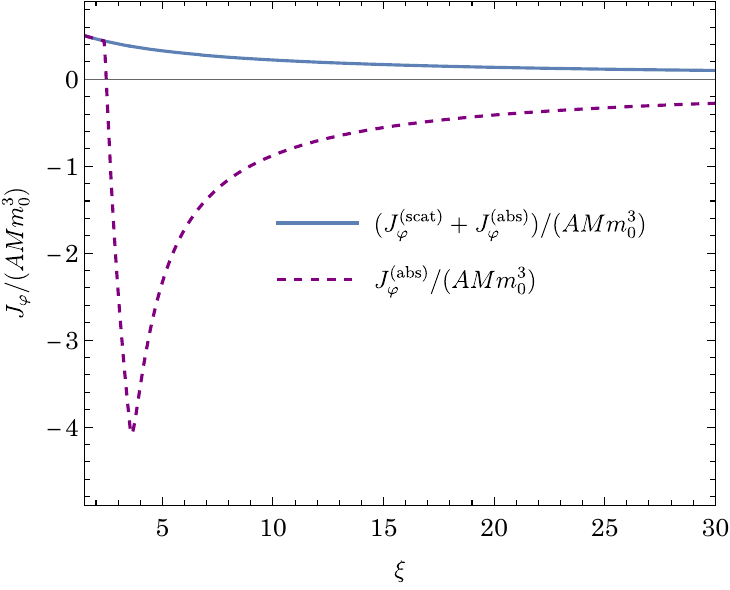}}
\caption{\label{fig:MJExamples} Sample graphs of $J_t/(A m_0^3)$ and $J_\varphi/(A M m_0^3)$ for the Maxwell-J\"{u}ttner model with $\alpha = 1/2$, $\beta = 1$, $\xi_0 = 30$. The components corresponding to absorbed trajectories are plotted with dashed lines. Solid lines represent the sums $(J_t^\mathrm{(abs)} + J_t^\mathrm{(scat)})/(A m_0^3)$ and $(J_\varphi^\mathrm{(abs)} + J_\varphi^\mathrm{(scat)})/(A M m_0^3)$.}
\end{figure}

Sample graphs of the components $J_t/(A m_0^3)$ and $J_\varphi/(A m_0^3)$ are shown in Fig.\ \ref{fig:MJExamples}. As in Figs.\ \ref{fig:jtmono} and \ref{fig:jphimono}, we plotted components corresponding to absorbed trajectories, as well as sums $(J_t^\mathrm{(abs)} + J_t^\mathrm{(scat)})/(A m_0^3)$ and $(J_\varphi^\mathrm{(abs)} + J_\varphi^\mathrm{(scat)})/(A m_0^3)$. We choose $\Lambda_\mathrm{max}(\xi,\xi_0,\varepsilon,\epsilon_\sigma)$ defined by Eq.\ (\ref{Lambdamax}). In principle, one could expect some sort of non-smooth behavior shown in Figs.\ \ref{fig:jtmonomin} and \ref{fig:jphimonomin} for fine-tuned energies to be present also in the Maxwell-J\"{u}ttner model. Is seems that it is smeared by contributions from other energies.

\subsection{Accretion rates}

\begin{figure}[t]
\subfigure[]{
\includegraphics[width=0.49\textwidth]{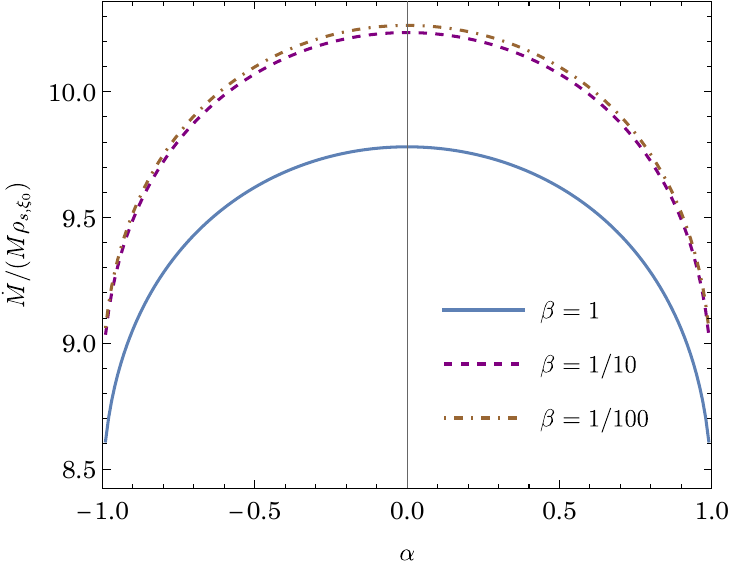}}
\subfigure[]{
\includegraphics[width=0.49\textwidth]{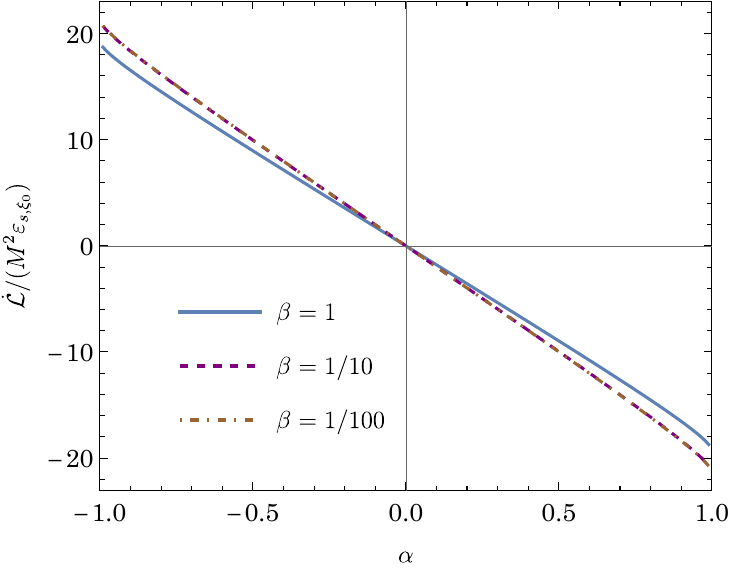}}
\caption{\label{fig:MdotLdotJuttner} Rest mass and angular momentum accretion rates for Maxwell-J\"{u}ttner type distribution  models with $\xi_0 = 30$. We plot results for $\beta = 1$, 1/10 and 1/100.}
\end{figure}

Assuming Eq.\ (\ref{MJdist}), we get the following expressions for accretion rates
\begin{subequations}
\label{accrratesMJ}
\begin{eqnarray}
\dot M & = & 2 \pi A M m_0^4 \sum_{\epsilon_\sigma = \pm 1}  \left\{ \int_{W_\mathrm{min}(\xi_0,\epsilon_\sigma)}^{\varepsilon_\mathrm{crit}(\xi_0,\epsilon_\sigma)} d\varepsilon e^{-\beta \varepsilon} \left[\lambda_\mathrm{max}(\xi_0,\varepsilon,\epsilon_\sigma)-\Lambda_\mathrm{min}(\xi_0,\varepsilon,\epsilon_\sigma)\right] +  \int_{\varepsilon_\mathrm{crit}(\xi_0,\epsilon_\sigma)}^\infty d \varepsilon e^{-\beta \varepsilon} \lambda_c(\varepsilon,\epsilon_\sigma) \right\}, \\
\dot {\mathcal E} & = & 2 \pi A M m_0^4 \sum_{\epsilon_\sigma = \pm 1}  \left\{ \int_{W_\mathrm{min}(\xi_0,\epsilon_\sigma)}^{\varepsilon_\mathrm{crit}(\xi_0,\epsilon_\sigma)} d\varepsilon \varepsilon e^{- \beta \varepsilon} \left[\lambda_\mathrm{max}(\xi_0,\varepsilon,\epsilon_\sigma)-\Lambda_\mathrm{min}(\xi_0,\varepsilon,\epsilon_\sigma)\right] +  \int_{\varepsilon_\mathrm{crit}(\xi_0,\epsilon_\sigma)}^\infty d \varepsilon \varepsilon e^{- \beta \varepsilon} \lambda_c(\varepsilon,\epsilon_\sigma) \right\}, \\
\dot {\mathcal L} & = & 2 \pi A M^2 m_0^4 \nonumber \\
&& \times \sum_{\epsilon_\sigma = \pm 1}  \left\{ \int_{W_\mathrm{min}(\xi_0,\epsilon_\sigma)}^{\varepsilon_\mathrm{crit}(\xi_0,\epsilon_\sigma)} d\varepsilon e^{- \beta \varepsilon} \left[ \frac{\epsilon_{\sigma}}{2} \big[\lambda^2_\mathrm{max}(\xi_0,\varepsilon,\epsilon_\sigma)-\Lambda^2_\mathrm{min}(\xi_0,\varepsilon,\epsilon_\sigma)\big] +\alpha \varepsilon\big[\lambda_\mathrm{max}(\xi_0,\varepsilon,\epsilon_\sigma)-\Lambda_\mathrm{min}(\xi_0,\varepsilon,\epsilon_\sigma) \big] \right]
\right. \nonumber \\
&& \left. +  \int_{\varepsilon_\mathrm{crit}(\xi_0,\epsilon_\sigma)}^\infty d \varepsilon e^{-\beta \varepsilon} \left[ \frac{\epsilon_\sigma}{2} \lambda_c(\varepsilon,\epsilon_\sigma) + \alpha \varepsilon \right] \lambda_c(\varepsilon,\epsilon_\sigma) \right\}.
\end{eqnarray}
\end{subequations}

Sample graphs of $\dot M$ and $\dot{\mathcal L}$ obtained from Eqs.\ (\ref{accrratesMJ}) are shown in Fig.\ \ref{fig:MdotLdotJuttner}. We assume different values of $\beta$ and plot the dependence of $\dot M$ and $\dot{\mathcal L}$ on the black hole spin parameter $\alpha$. Both plots exhibit a qualitative behavior similar to the one obtained in \cite{CMO2022}. Also note that the curious features appearing for fine-tuned energies in the monoenergetic model illustrated in Fig.\ \ref{fig:MdotLdotmono} are no longer present in Fig.\ \ref{fig:MdotLdotJuttner} (they too become smeared by contributions corresponding to different energies).

\begin{figure}[t]
\subfigure[]{\label{mdot}\includegraphics[width=0.49\textwidth]{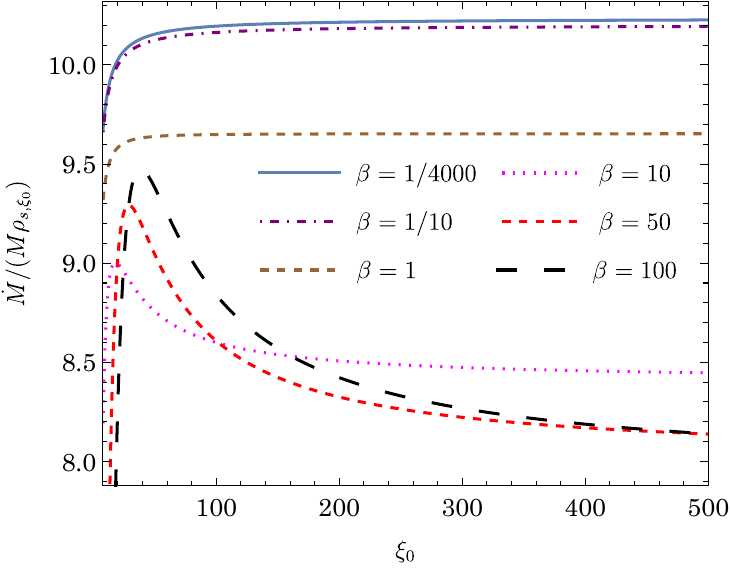}}
\subfigure[]{\label{edot}\includegraphics[width=0.49\textwidth]{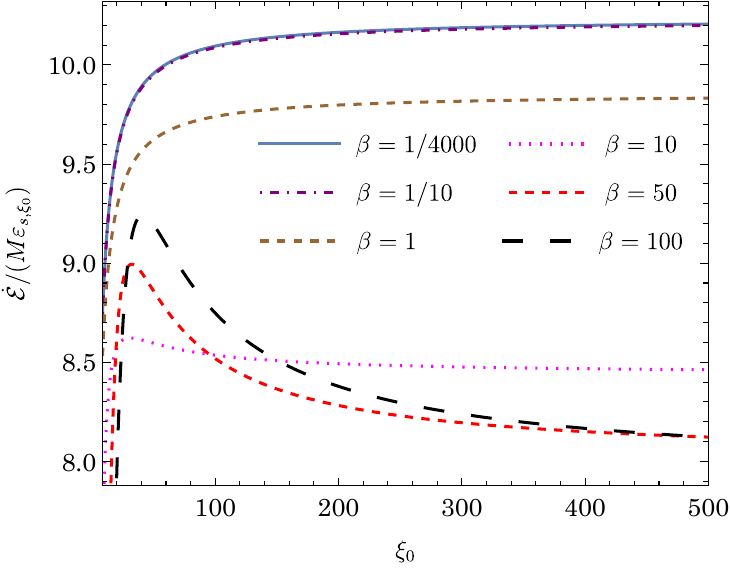}}
\subfigure[]{\label{ldot}\includegraphics[width=0.49\textwidth]{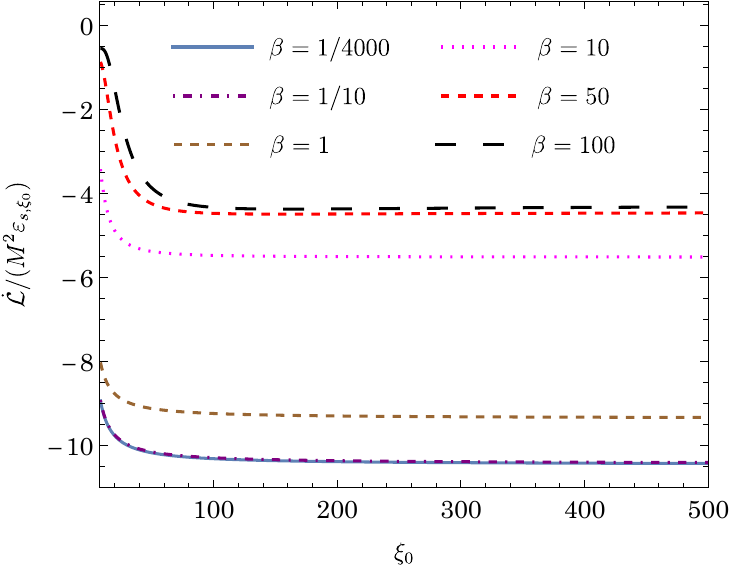}}
\caption{\label{fig:ratesvsxi0} Accretion rates for Maxwell-J\"{u}ttner type distribution  models vs finite radius $\xi_0$. \ref{mdot} Rest mass accretion rates. \ref{edot} Energy accretion rates. \ref{ldot} Angular momentum accretion rates. In all three cases, the value of $\alpha$ is consistently set to 0.5.}
\end{figure}

Figure \ref{fig:ratesvsxi0} illustrates the dependence of accretion rates $\dot M$, $\dot{\mathcal E}$, and $\dot{\mathcal L}$ on $\xi_0$. We plot the data for the range $\xi_0 \in [10,500]$, and several different values of $\beta \in [1/4000,100]$. The changes of the accretion rates with $\xi_0$ and $\beta$ seem to be of comparable magnitude. On one hand, this means that in precise applications the location of $\xi_0$ has to be taken into account. On the other, for rough estimates, one can rely on the values provided by a simple toy model, in which $\xi_0$ is infinite \cite{CMO2022}. The effects caused by choosing different values of $\xi_0$ are most pronounced for small $\xi_0$. Our limit $\xi_0 > 10$ was chosen for convenience, assuring that $\xi_0$ was always larger than the location of the marginally stable circular orbit $\xi_\mathrm{ms}$ (the values given by Eq.\ (\ref{xims}) for $-1 \le \alpha \le 1$ satisfy $\xi_\mathrm{ms} \le 9$, with the equality for $\epsilon_\sigma \alpha = -1$). 

The limiting values of $\dot M$, $\dot{\mathcal E}$, and $\dot{\mathcal L}$ for $\xi_0 \to \infty$ were computed in \cite{CMO2022}. Assuming $\xi_0 = \infty$, we get, in particular, the following limits for $\beta \to 0$ and $\beta \to \infty$:
\begin{subequations}
\begin{eqnarray}
\lim_{\beta \to \infty} \dot{M} & = &  2 M \rho_{s,\infty} \left( 2+\sqrt{1-\alpha} + \sqrt{1+\alpha} \right), \\
\lim_{\beta \to \infty} \dot{\mathcal{E}} & = & 2 M \rho_{s,\infty} \left( 2 + \sqrt{1 - \alpha} + \sqrt{1 + \alpha} \right), \\
\lim_{\beta \to \infty} \dot{\mathcal{L}} & = & 4 M^2 \rho_{s,\infty} \left( \sqrt{1 - \alpha} - \alpha - \sqrt{1 + \alpha} \right), \\
\lim_{\beta \to 0} \dot{M}  & = &  6 \sqrt{3} M \rho_{s,\infty} \cos \left( \frac{1}{3} \arcsin{\alpha} \right), \\
\lim_{\beta \to 0}\dot{\mathcal{E}} & = & 6 \sqrt{3} M \varepsilon_{s,\infty} \cos \left(\frac{1}{3} \arcsin{\alpha}\right), \\
\lim_{\beta \to 0}\dot{\mathcal{L}} & = &  -6 \sqrt{3} M^2 \varepsilon_{s,\infty} \sin \left(\frac{2 \arcsin{\alpha} }{3}\right) \left[ 2+\cos \left(\frac{2 \arcsin{\alpha} }{3}\right)\right],
\end{eqnarray}
\end{subequations}
where $\rho_{s,\infty}$ and $\varepsilon_{s,\infty}$ denote the asymptotic values of the surface rest-mass and energy densities, respectively. Assuming, as in Fig.\ \ref{fig:ratesvsxi0}, $\alpha = 1/2$, one obtains
\begin{equation}
\label{limits12a}
    \lim_{\beta \to \infty} \dot{M} = 7.8637 M \rho_{s,\infty}, \quad \lim_{\beta \to \infty} \dot{\mathcal{E}} = 7.8637 M \rho_{s,\infty}, \quad \lim_{\beta \to \infty} \dot{\mathcal{L}} = -4.07055 M^2 \rho_{s,\infty},
\end{equation}
and
\begin{equation}
\label{limits12b}
    \lim_{\beta \to 0} \dot{M} = 10.2344 M \rho_{s,\infty}, \quad \lim_{\beta \to 0} \dot{\mathcal{E}} = 10.2344 M \varepsilon_{s,\infty}, \quad \lim_{\beta \to 0} \dot{\mathcal{L}} = -10.4488 M^2 \varepsilon_{s,\infty}.
\end{equation}
Also note that for $\beta \to \infty$ (vanishing asymptotic temperature), we have $\rho_{s,\infty} \to \varepsilon_{s,\infty}$. The values given in Eqs.\ (\ref{limits12a}) and (\ref{limits12b}) can be compared with the data in Fig.\ \ref{fig:ratesvsxi0}.

\section{Summary}
\label{sec:summary}

We investigated stationary general-relativistic kinetic models of razor-thin accretion disks confined to the equatorial plane of the Kerr spacetime. Our models aim to improve a previous toy-model of a disk extending to infinity within the equatorial plane \cite{CMO2022}.

For simplicity, we only considered a gas of non-colliding same-mass spinless particles. Consequently, our model could be applied directly to modeling of dark matter, in the spirit of Refs.\ \cite{pmao1,CMO2022,Dominguez2017}. However, our motivation to tackle this topic was different and was, in fact, twofold. On one hand, we wanted to remove the slightly artificial assumption present in the original analysis of \cite{CMO2022} that the disk extends to infinity. On the other hand, the kinetic model with a boundary condition placed at a finite distance from the Kerr black hole provides an ideal setting to understand the mechanics of the geodesic motion in the Kerr spacetime from the phase-space perspective. We would expect similar effects to be present also for models including magnetic fields and, to some extent, models taking into account the self-gravity of the gas (Einstein-Vlasov systems).

Regarding more explicit physical results, one should state that the change of the location of the outer edge of the disk affects the obtained accretion rates in a similar order of magnitude as the asymptotic (or boundary) temperature of the gas. This implies, in particular, that very general estimates based on the infinite model of \cite{CMO2022}---such as an estimate of a characteristic time scale in which a spinning black hole slows down its rotation due to accretion of dark matter---should remain valid. Of course, in more precise estimates, the location of the outer disk boundary has to be taken into account on equal footing with other asymptotic parameters of the gas.

\begin{acknowledgments}
P.\ M.\ acknowledges a support of the Polish National Science Centre Grant No.\ 2017/26/A/ST2/00530.
\end{acknowledgments}

\appendix
\section{Angular momentum integrals}

All integrals with respect to $\lambda$ appearing in Eqs.\ (\ref{e0abs}) and (\ref{e0scat}) can be computed analytically. Note that $\tilde R = -\xi (\xi - 2)(\lambda - \lambda_1)(\lambda - \lambda_2)$. This gives, for $\xi > 0$, $\tilde R = \xi (\xi-2)(\lambda - \lambda_1)(\lambda_2 - \lambda)$. It is easy to check that
\begin{eqnarray}
    \int_{\lambda_1}^\lambda \frac{d \lambda^\prime}{\sqrt{(\lambda^\prime - \lambda_1)(\lambda_2 - \lambda^\prime)}} & = & - 2 \arctan \sqrt{ \frac{\lambda_2 - \lambda}{\lambda - \lambda_1}}, \\
    \int_{\lambda_1}^\lambda \frac{ \lambda^\prime d \lambda^\prime}{\sqrt{(\lambda^\prime - \lambda_1)(\lambda_2 - \lambda^\prime)}} & = & - \sqrt{(\lambda - \lambda_1)(\lambda_2 - \lambda)} + (\lambda_1 + \lambda_2) \arcsin \sqrt{\frac{\lambda - \lambda_1}{\lambda_2 - \lambda_1}},
\end{eqnarray}
where $\lambda_1 < \lambda < \lambda_2$. The appropriate expressions for the definite integrals appearing in Eqs.\ (\ref{e0abs}) and (\ref{e0scat}) follow immediately from these formulas. For $0 < \xi < 2$, we have $- \xi (\xi - 2) > 0$. In this case we are interested in the integrals of the form
\begin{eqnarray}
    \int_{\lambda_2}^\lambda \frac{d \lambda^\prime}{\sqrt{(\lambda^\prime - \lambda_1)(\lambda^\prime - \lambda_2)}} & = & - \ln \left( 1 - \sqrt{\frac{\lambda - \lambda_2}{\lambda - \lambda_1}} \right) + \ln \left( 1 + \sqrt{\frac{\lambda - \lambda_2}{\lambda - \lambda_1}} \right), \\
    \int_{\lambda_2}^\lambda \frac{\lambda^\prime d \lambda^\prime}{\sqrt{(\lambda^\prime - \lambda_1)(\lambda^\prime - \lambda_2)}} & = & \sqrt{(\lambda - \lambda_1)(\lambda - \lambda_2)} + (\lambda_1 + \lambda_2) \mathrm{arsinh} \sqrt{\frac{\lambda - \lambda_2}{\lambda_2 - \lambda_1}},
\end{eqnarray}
for $\lambda_1 < \lambda_2 < \lambda$. The appropriate forms for $\lambda < \lambda_1 < \lambda_2$ read
\begin{eqnarray}
    \int_\lambda^{\lambda_1} \frac{d \lambda^\prime}{\sqrt{(\lambda^\prime - \lambda_1)(\lambda^\prime - \lambda_2)}} & = & \ln \left( 1 + \sqrt{\frac{\lambda_2 - \lambda}{\lambda_1 - \lambda}} \right) - \ln \left( - 1 + \sqrt{\frac{\lambda_2 - \lambda}{\lambda_1 - \lambda}} \right), \\
    \int_\lambda^{\lambda_1} \frac{\lambda^\prime d \lambda^\prime}{\sqrt{(\lambda^\prime - \lambda_1)(\lambda^\prime - \lambda_2)}} & = & - \sqrt{(\lambda - \lambda_1)(\lambda - \lambda_2)} + (\lambda_1 + \lambda_2) \mathrm{artanh} \sqrt{\frac{\lambda_1 - \lambda}{\lambda_2 - \lambda}}.
\end{eqnarray}
For $\xi < 2$, one may also encounter a case in which $\tilde R$ is strictly positive, i.e., without real zeros $\lambda_1$ and $\lambda_2$. One can then use the form
\begin{equation}
    \int \frac{d \lambda}{\sqrt{a \lambda^2 + b \lambda + c}} = - \frac{1}{\sqrt{a}} \ln \left| b + 2 a \lambda - 2 \sqrt{a} \sqrt{a \lambda^2 + b \lambda + c} \right| + \mathrm{const}.
\end{equation}

\end{document}